\documentclass[12pt]{article}

\usepackage[T1]{fontenc} 				

\usepackage{amssymb,amsmath,amscd,extarrows}
\usepackage[english]{babel}
\usepackage{titlesec}
\usepackage{slashed}
\usepackage{hyperref}
\usepackage{graphicx,xcolor}
\usepackage{enumerate}
\usepackage{bbm} 						
\usepackage{float}
\usepackage{mathtools}
\usepackage{subfig}
\usepackage{setspace} 

\hypersetup{
	colorlinks=true,
	linkcolor=black,
	citecolor=dark-red,
	urlcolor=dark-green,
	linktoc=all
}

\usepackage{geometry} 					
\numberwithin{equation}{section} 

\titleformat{\section}[block]{\Large\bfseries\centering}{\thesection}{1em}{} 
\titleformat{\subsection}[block]{\bfseries}{\thesubsection}{1em}{} 

\definecolor{dark-gray}{gray}{0.20}
\definecolor{gray}{gray}{0.30}
\definecolor{light-gray}{gray}{0.80}
\definecolor{dark-red}{rgb}{0.7,0,0}
\definecolor{dark-green}{rgb}{0.1,0.4,0}
\definecolor{dark-blue}{rgb}{0.3,0.3,0.7}
\definecolor{light-blue}{rgb}{0.8,0.8,1}
\definecolor{cardinal}{rgb}{0.6,0,0}
\definecolor{darkgreen}{rgb}{0,0.5,0}
\definecolor{golden}{rgb}{0.92, 0.7, 0}
\definecolor{midnight}{rgb}{0, 0, 0.5}
\definecolor{darkblue}{rgb}{0.2, 0, 0.8}
\definecolor{forestgreen}{rgb}{0.13, 0.55, 0.13}

\def\cB{{\cal B}}
\def\cD{{\cal D}}
\def\cF{{\cal F}}

\def\cI{{\cal I}}

\def\cM{{\cal M}}

\def\cR{{\cal R}}

\newcommand{\cE}{\mathcal{E}}

\newcommand{\bea}{\begin{eqnarray}}
\newcommand{\eea}{\end{eqnarray}}
\newcommand{\be}{\begin{equation}}
\newcommand{\ee}{\end{equation}}

\title{\bf \fontsize{20pt}{24pt}{Partition functions on squashed seven-spheres and holography\\
	}\vspace{3mm}}

\bigskip
\bigskip

\author{ Xuao Zhang\\[5mm]
	\normalsize Instituut voor Theoretische Fysica, K.U. Leuven\\
	\normalsize Celestijnenlaan 200D, BE-3001 Leuven, Belgium\\[2mm]
	\texttt{\small  \href{mailto::xuao.zhang@kuleuven.be}{xuao.zhang@kuleuven.be}}\\
}

\date{}

\begin{document}  
	
\maketitle
\begin{abstract}
	\noindent Our paper presents two main results. First, we study the renormalized free energies of Euclidean Einstein gravity in asymptotically AdS$_8$ and various field theories on a squashed seven sphere. In the gravity theory, we demonstrate the absence of the Hawking-Page transition, while in the field theory, we focus on the O($N$) vector model and the massless free fermion model. The conformal symmetry governs the universal behaviors of the free energies for small and large squashings, which we confirm numerically and analytically. Second, we evaluate the second-order derivative of CFT free energy with respect to the squashing parameter, finding universal results that hold for generic conformal field theories. We examine two different squashings, one with an SU(2) bundle, which is the primary focus of our paper, and another with a U(1) bundle, where our results align with the conjectured formula from the gravity side in the literature.
\end{abstract}

\clearpage

\setlength{\parskip}{1pt}
\tableofcontents
\setlength{\parskip}{1em}

\clearpage
\newpage

\section{Introduction}

The AdS/CFT correspondence posits dualities between large $N$ field theories and string theories in asymptotically AdS spacetimes. \cite{Maldacena:1997re, Witten:1998qj, Gubser:1998bc} In its original proposal, the conformal field theory is living on a manifold $\cM_d$ conformally equivalent to the boundary of the bulk. The coupling between the conformal field theory and the background helps us gain insight into the dynamics of the field theory. As the bulk theories typically lie in the semi-classical regime, one usually needs to exert non-perturbative tools to investigate the field theories, the well-known examples include supersymmetric localization \cite{Pestun:2016zxk}, bootstrap, and integrability. Through a simplified version of AdS/CFT correspondence \cite{Klebanov:2002ja}, the O($N$) model, which is a set of $N$ conformally coupled scalars, corresponds to higher spin gravity in the bulk, both of them can be attacked analytically. In our setup, instead, we consider a Euclidean Einstein gravity living in an asymptotically locally AdS$_8$ space without matter coupling, which does not correspond to O($N$) model, but shares some universal behaviors as dictated by the conformal symmetry. On the boundary, we consider non-supersymmetric conformal field theories.

The phase structure of asymptotically AdS spaces is an interesting topic, originating from the famous Hawking-Page phase transition \cite{Hawking:1982dh} between thermal AdS$_4$ space and Euclidean Schwarzschild black hole. Holographically, the Hawking-Page transition is conjectured to be dual to the confinement-deconfinement transition in QCD, where the two phases correspond to different behaviors of Wilson loops. \cite{Witten:1998zw} Generalizations of Hawking-Page transition has been discussed in \cite{Blackman:2011in} where the asymptotical boundary geometry is $S^2\times S^d$, and in \cite{Aharony:2019vgs} where the boundary geometry is $S^{d_1} \times S^{d_2}$. Besides, thermodynamical properties and phase transitions are also discussed for AdS-Taub-NUT spaces \cite{Khodam-Mohammadi:2008hww, Chamblin:1998pz, Awad:2000gg, Clarkson:2002uj, Astefanesei:2004kn}, where the asymptotic geometry is a U(1) fiber bundle over a K\"ahler-Einstein space $\cB$, where the metric looks like:
\be 
ds^2 = dr^2 + a^2(r) ds^2_{\cB} + b^2(r) (d\psi + A_\cB)^2,
\ee 
where $A_\cB$ is the K\"ahler potential and the K\"ahler form is given by $J_\cB = dA_\cB$. The metric on the asymptotic boundary where the conformal field theory lives is given by taking $r\rightarrow \infty$:
\be 
ds^2_{\rm bdy} = ds_\cB^2 + \lambda^2 (d\psi + A_\cB)^2,\quad \lambda \equiv \lim\limits_{r\rightarrow \infty} \frac{b(r)}{a(r)}.
\ee 
The squashing parameter $\lambda$ controls the size of the U(1) bundle, which no longer has the interpretation as the inverse temperature as in thermal field theories. Besides NUT, there exists another family of spaces, which has the same asymptotic symmetry but with a different topology, dubbed Bolt, which has a horizon where the Killing vectors degenerate. By varying the size of the U(1) bundle and comparing the gravitational free energies, one observes a first-order phase transition between Taub-NUT and Taub-Bolt. In \cite{Bobev:2016sap}, a generalization of Euclidean AdS$_4$-Taub-NUT space with two squashing parameters was studied.

In our work, we focus on a squashed asymptotically locally AdS$_8$ geometry preserving SO(5)$\times$SO(3) isometry \cite{Page:1985hg,Gibbons:1989er}\footnote{See \cite{Hiragane:2003qq} for a similar bulk setup but with positive cosmological constant. }, whose boundary geometry is a squashed seven sphere constructed by an SU(2) bundle over $\mathbb{S}^4$. By evaluating and comparing the gravitational free energy of SU(2) analog of NUT and Bolt spaces, we are able to study the phase structure of our AlAdS$_8$ geometry. One special property of our squashed $\mathbb{S}^7$ is that there exist two squashing parameters $\lambda = 1, \lambda^2 = 1/5$, for which the squashed sphere is an Einstein manifold \cite{jensen1973einstein}, i.e., $R_{\mu\nu} = k g_{\mu\nu}$ for some constant $k$. For the two special cases, we are able to solve the Einstein equations analytically and then evaluate the Euclidean free energy by using the standard holographic renormalization. \cite{Clarkson:2002uj, Emparan:1999pm, Bueno:2022log} For general $\lambda > 0$, we can evaluate the free energy numerically. Interestingly, although having the analogs of NUT and Bolt phases, there exists only one phase for a given value $\lambda$, meaning there is no Hawking-Page-like transition between the two competing phases. A similar phase structure has been found in \cite{Aharony:2019vgs}, which appears only if the the boundary geometry $S^{d_1}\times S^{d_2}$ satisfies $d_1 + d_2 \ge 9$.

The field theory contents we're interested in are Euclidean conformal field theories living on the squashed seven sphere with SO(5)$\times$SO(3) isometry, where the squashing naturally couples the field theory with the background metric. It is the stress tensor of boundary CFT that couples to the squashed metric, the dynamical properties of which are constrained largely by the conformal symmetry \cite{Osborn:1993cr,Erdmenger:1996yc}, which reflects the universal character for all CFTs. The free energy in odd-dimensional field theories is an important quantity since it's conjectured to reflect the degrees of freedom in the field theory and thus monotonically decreasing along with RG flows, which is called $F$-theorem. \cite{Jafferis:2011zi, Pufu:2016zxm} The quantity conjectured to be decreasing along the RG flow is related to the free energy $F \equiv - \log |Z_{S^d}|$ by a sign:
\be 
\widetilde F \equiv (-1)^{\frac{d-1}{2}} \log |Z_{S^d}|.
\label{FreeEnergySd}
\ee 
The $F$-theorem was proven only in 3d CFT by \cite{Casini:2012ei}, but is supported by pieces of evidence in other dimensions. \cite{Pufu:2016zxm} In our scenario, the CFT living on round $\mathbb{S}^7$ is a free UV fixed point, and introducing squashing on the metric corresponds to switching on a marginal spin-2 deformation generated by the stress tensor: \cite{Bobev:2017asb}
\be 
S[g_{\mu\nu}, \phi] = S[g^{(0)}_{\mu\nu},\phi] - \frac{\epsilon}{2} \int d^d x \sqrt{g^{(0)}} h^{\mu\nu}T_{\mu\nu},\quad T_{\mu\nu} \equiv -\frac{2}{\sqrt{g}} \frac{\delta S[g_{\mu\nu},\phi]}{\delta g^{\mu\nu}} .
\label{Deformation}
\ee 
For a conformal field theory without conformal anomaly, which is true in odd dimensions, we expect the first-order derivative $\widetilde F'(\epsilon)$ to vanish at $\epsilon = 0$, which is proportional to the one-point function of the stress tensor. And the second-order derivative $\widetilde F''(\epsilon)$ should be negative as dictated by $F$-theorem since the marginal spin-2 deformation induces an RG flow from the UV fixed point, along which the free energy must be decreasing. It can be expressed as a double integral on the round sphere:
\be
F''(\epsilon)\Big|_{\epsilon=0} = -\frac{1}{4}\int d^dx d^dy\sqrt{g^{(0)}(x)g^{(0)}(y)} h^{ab}(x)h^{cd}(y)\langle T_{ab}(x)T_{cd}(y)\rangle_{\partial \cM}.
\ee
Quite remarkably, we are able to evaluate the second-order derivatives for general CFTs living on not only squashed SO(5)$\times$SO(3) seven spheres but also another U($k+1$)$\times$U(1) family of squashed $(2k+1)$-spheres, the latter was only conjectured from results in high-derivative gravity in \cite{Bueno:2018yzo}. The result is universal and applies to all CFTs, and this is the first time to obtain them from the field theory to our knowledge.

 The above universal results only apply to cases where the squashing is small. For finite squashing, we consider two toy models: the O($N$) vector model, which is equivalent to conformally coupled scalars, and a free fermion model. The free energy of the field theories can be obtained by taking advantage of the spectrum of the Laplacian and the Dirac operators on our squashed metric \cite{Nilsson:1983ru, Eastaugh:1985ew}, where the path integral boils down to a Gaussian integral, making the evaluation of the free energy for generic squashing $\lambda$ possible. \cite{Hartnoll:2005yc, Anninos:2012ft, Anninos:2013rza} We have also investigated the behavior of free energy at small squashing analytically and at large squashing numerically. The comparison among free energies of holographic CFTs, the O($N$) model, and the free fermion model has been studied at small squashing \cite{Bobev:2017asb, Bueno:2018yzo, Bueno:2020odt} and large squashing \cite{Bobev:2017asb, Hartnoll:2005yc, Anninos:2012ft} where the boundary is a squashed three-sphere. We have extended this exploration to the seven-sphere with SU(2) bundle.

The structure of the paper is as follows. In Section \ref{BulkTheory}, we introduce two bulk metric ansatzes which preserve SO(5)$\times$SO(3) isometry and solve the Einstein equations. We also calculate the gravitational free energy as a function of the squashing parameter $\lambda$. The field theory calculations are included in Section \ref{BoundaryTheory}, where we begin with the small-squashing behavior for generic CFT living on SU(2) squashed seven spheres and U(1) squashed $(2k+1)$-spheres. We also evaluate the free energy of both conformally-coupled scalar and free fermion fields using different methods. In section \ref{SectionofComparison}, we compare bulk and boundary free energies.

\section{Bulk story}
\label{BulkTheory}
In this section, we study the generalization of Hawking-Page transition in asymptotically locally AdS$_8$ with Euclidean signature. We begin with introducing two ways to construct squashed sphere metrics and use them to obtain the metrics we are after. Taking the squashed seven sphere as a boundary, the bulk geometry can be obtained by numerically solving the equation of motions. In the end, we can take advantage of the bulk solution to obtain the renormalized free energy.

\subsection{A tale of two metrics}\label{ATableTwoMetrics}
In this paper, we consider squashed seven spheres with SO(5)$\times$SO(3) isometry, whose metric can be constructed by distinguished yet equivalent ways. The first one is by embedding the sphere in a projective space admitting the standard Fubini-Study metric, and the other one is by Hopf fibration over a projective space, the fibration is non-trivial because of the existence of K\"ahler or hyper-K\"ahler potential on the projective space.

We start with the first construction following \cite{Awada:1982pk,Duff:1986hr}, where we embed $\mathbb{S}^7$ in $\mathbb{HP}^2$. The standard Fubini-Study metric on $\mathbb{HP}^2$ is:
\be 
ds^2 = (1+\bar{q}_kq_k)^{-1} d\bar{q}_i dq_i - (1+\bar{q}_kq_k)^{-2}\bar{q}_idq_id\bar{q}_jq_j,\quad q_1,q_2\in \mathbb{HP}^2,
\label{FubiniStudy}
\ee
where the repeated indices are summed over $\{1,2\}$. We introduce a parametrization of $(q_1, q_2)\in \mathbb{HP}^2$:
\be
q_1 = U\tan\chi \cos\frac{1}{2}\mu,\quad q_2 = V\tan\chi\sin\frac{1}{2}\mu,\quad 0\le \chi\le\pi/2,\quad 0\le \mu\le\pi.
\ee
Here, $U, V\in$ SU(2) can be regarded as quaternions with unit length, so we parametrize them by Euler angles $(\theta, \phi, \psi)$ and $(\Theta, \Phi, \Psi)$ respectively:
\be
U = e^{\textbf{k}\phi/2} e^{\textbf{i}\theta/2} e^{\textbf{k}\psi/2},\quad 
V = e^{\textbf{k}\Phi/2} e^{\textbf{i}\Theta/2} e^{\textbf{k}\Psi/2};\quad 0\le \theta\le \pi,\ 0\le \phi\le 2\pi,\ 0\le \psi \le 4\pi,
\label{parametrization}
\ee
where $\mathbf{i,j,k}$ are units of quaternion. The Maurer-Cartan form can be worked out directly:
\be
\begin{aligned}
	& 2U^{-1}dU = \mathbf{i}\sigma_1 + \mathbf{j}\sigma_2 + \mathbf{k}\sigma_3,\quad 2V^{-1}dV = \mathbf{i}\Sigma_1 +\mathbf{j}\Sigma_2 +\mathbf{k}\Sigma_3,	\\
\end{aligned}
\ee
where the left-invariant one-forms are:
\be \small \begin{aligned}
	& \sigma_1 = \cos\psi d\theta + \sin\psi\sin\theta d\phi,\quad \sigma_2 = -\sin\psi d\theta + \cos\psi\sin\theta d\phi,\quad \sigma_3 = d\psi + \cos\theta d\phi;\\
	& \Sigma_1 = \cos\Psi d\Theta + \sin\Psi\sin\Theta d\Phi,\quad \Sigma_2 = -\sin\Psi d\Theta + \cos\Psi\sin\Theta d\Phi,\quad \Sigma_3 = d\Psi + \cos\Theta d\Phi.\\
\end{aligned}\ee
Using the parametrization above, the metric (\ref{FubiniStudy}) becomes
\be\begin{aligned}
	ds^2 &= d\chi^2 + \frac{1}{4}\sin^2\chi\left[d\mu^2 + \frac{1}{4}\sin^2\mu \omega_i^2 + \frac{1}{4}\cos^2\chi(\nu_i + \cos\mu \omega_i)^2\right],\\
	& \quad \quad \nu_i \equiv \sigma_i + \Sigma_i,\quad  \omega_i \equiv \sigma_i - \Sigma_i,\quad i = 1,2,3.\\
\end{aligned}
\ee
By setting $\chi$ to be constant, we're equivalently taking the co-dimension 1 surface foliating the projective space and surrounding the original point $\chi=0$. Introducing the squashing parameter $\lambda \equiv \cos\chi$, we obtain the metric on a squashed seven sphere with unit radius \cite{Awada:1982pk}:\footnote{The scale factor $1/4$ ahead is to make sure when $\lambda = 1$, the metric goes back to unit sphere $\mathbb{S}^7$. }
\be
ds^2 = \frac{1}{4} \left[ d\mu^2 +\frac{1}{4}\sin^2\mu\omega_i^2 + \frac{1}{4}\lambda^2 (\nu_i +\cos\mu \omega_i)^2 \right].
\label{SquashedSphereMetric1}
\ee
One can introduce the following vielbein:
\be 
e^1 = \frac{1}{2}d\mu,\quad e^i = \frac{1}{4}\sin\mu\omega_{i-1},\quad e^I = \frac{\lambda}{4}(\nu_{I-4} + \cos\mu \omega_{I-4}),\quad i = 2,3,4;\ I = 5,6,7.
\ee 
Under the tetrad, the Ricci tensor is diagonal:
\be 
R_{ab} = {\rm diag} \left(\alpha,\alpha,\alpha,\alpha,\beta,\beta,\beta\right),\quad \alpha = 12-6\lambda^2,\quad \beta = 4\lambda^2 + \frac{2}{\lambda^2}.
\label{RicciTensor}
\ee 
For $\alpha = \beta$, the space becomes Einstein. There're two possibilities for that condition, $\lambda = 1$ and $\lambda_* = \frac{1}{\sqrt{5}}$. These two values will both play important roles in the gravitational free energy.

Now let's look at the other way to construct squashed seven-sphere with $SO(5)\times SO(3)$ isometry. It's motivated by $k=1$ SU(2) Yang-Mills instanton on $\mathbb{S}^4$. Considering $\mathbb{S}^7$ as $\mathbb{S}^3$ bundle over $\mathbb{S}^4$ with a gauge potential $A_i = \cos^2\frac{\tilde{\mu}}{2}\tilde{\Sigma}_i$, the metric is
\be
ds^2 = \frac{1}{4}\left[ds^2_{\mathbb{S}^4} + \lambda^2 (\tilde{\sigma}_i - A_i)^2 \right],\quad ds^2_{\mathbb{S}^4} = d\tilde{\mu}^2 + \frac{1}{4}\sin^2\tilde{\mu}\tilde{\Sigma}_i^2.
\label{SquashedSphereMetric2}
\ee
This is an ``inverse-Kaluza-Klein'' procession, where we construct a metric on $\mathbb{S}^7$ out of a four-dimensional metric plus an SU(2) bundle, and the size of the bundle is described by the squashing parameter $\lambda$. In fact, we find the metric returns to $\mathbb{S}^3 \times \mathbb{S}^4$ if we set $A^i = 0$, where $\tilde{\sigma}_i$ parametrizes $\mathbb{S}^3$ and $\tilde{\Sigma}_i$ is on $\mathbb{S}^4$. One can define the following vielbein:
\be 
e^1 = \frac{1}{2} d\tilde{\mu},\quad e^i = \frac{1}{4}\sin\tilde{\mu}\tilde{\Sigma}_{i-1},\quad e^I = \frac{\lambda}{2}\left(\tilde{\sigma}_{I-4} - A_{I-4}\right),\quad i=2,3,4;\ I=5,6,7,
\ee 
under which the Ricci tensor is the same as the before \eqref{RicciTensor}, indicating that the two metrics are closely related.

The two metrics (\ref{SquashedSphereMetric1}) and (\ref{SquashedSphereMetric2}) constructed above share some similarities, they are both characterized by one squashing parameter $\lambda$, become round when $\lambda = 1$, and are Einstein when $\lambda = 1$ or $\frac{1}{\sqrt{5}}$; they both preserve explicitly the SO(5)$\times$SO(3) isometry. Although they're different metrics, there exists a map between them\footnote{The relation in equation (8.1.32) of \cite{Duff:1986hr} is not correct, and should actually be inversed.}
\be
\tilde{\sigma}_i = \sigma_i,\quad \mathbf{i}\Sigma_1 + \mathbf{j}\Sigma_2 + \mathbf{k}\Sigma_3 = \tilde{V}(\mathbf{i}\tilde{\omega}_1  + \mathbf{j}\tilde{\omega}_2 + \mathbf{k}\tilde{\omega}_3)\tilde{V}^{-1},\quad \tilde{\mu} = \pi - \mu,
\label{relation}
\ee
where $\tilde{\omega}_i \equiv \tilde{\sigma}_i - \tilde{\Sigma}_i$, and $2\tilde{V}^{-1}d\tilde{V} = \mathbf{i}\tilde{\Sigma}_1 + \mathbf{j}\tilde{\Sigma}_2 + \mathbf{k}\tilde{\Sigma}_3$ same as defined before. The map can be written equivalently in the following more compact way:
\be 
U = \tilde{U},\quad V = \tilde{U}\tilde{V}^{-1}\quad  \Leftrightarrow \quad \tilde{U} = U,\quad \tilde{V} = V^{-1}U;\quad \tilde{\mu} = \pi - \mu,
\label{Utilde}
\ee 
recall that $U, V$ are parametrized by the Euler angles as in \eqref{parametrization} and likewise for $\tilde{U}, \tilde{V}$. The relation above is nothing but a twist performed in the definition of some angles on the $\mathbb{S}^3$ bundle between the two metrics, which is philosophically the same as the transformation between (\ref{StdMetricS3}) and (\ref{MostStdMetricS3}) for a similar construction on complex projective spaces, this explains the similarities between the two metrics.

Since we're concerning Euclidean ${\rm AlAdS_8}$ in the bulk with a squashed seven sphere at the boundary, we make the following two ansatzes:\footnote{We have replaced all tilded quantities with un-tilded ones for simplicity.}
\begin{subequations}
	\begin{align}
		\label{Ansatz1}
		ds^2 &= dr^2 + a^2(r) \left(d\mu^2 + \frac{1}{4}\sin^2\mu \,\Sigma_i^2\right) + b^2(r) (\sigma_i - A_i)^2,\quad i = 1,2,3 \\
		ds^2 &= f_1^2(r)dr^2 + f_2^2(r)\left(d\mu^2 + \frac{1}{4}\sin^2\mu \omega_i^2 \right) +\frac{ f_{i+2}^2(r)}{4}(\nu_i + \cos\mu\omega_i)^2,  \label{Ansatz2}
\end{align} \end{subequations}
where $a(r),b(r),f_1(r),...,f_5(r)$ are undetermined functions of $r$ only. In the first ansatz, we set the coefficient of $dr$ to be 1 for simplicity, which can also be kept in general. Also note that in the second ansatz, $f_3, f_4,$ and $f_5$ can be different, while in the first ansatz, Einstein equations require only one function $b(r)$ can be arbitrary. This suggests an enhanced permutation symmetry in the ansatz (\ref{Ansatz1}), and the twist performed in \eqref{relation} breaks the symmetry. In the main text of the paper, we will focus on the first ansatz, and we put our partial results on the second ansatz in appendix \ref{solutionAnsatz1} and a more general study for future work.

\subsection{Equation of Motion}
We consider Einstein gravity in Euclidean AlAdS$_{d+1}$ space with negative cosmology constant, the equations of motion are:
\be 
G_{\mu\nu} + \Lambda g_{\mu\nu}=0 ,\quad \Lambda = -\frac{d(d-1)}{2\ell^2},
\ee 
where $\ell$ is the scale of AdS. Plugging in the ansatz, we take advantage of the vielbein formalism in GR to simplify the Einstein field equations. We take the achtbein $e^m_\mu$ such that the metric tensor $g_{\mu\nu} = \delta_{mn}e^m_\mu e^n_\nu$. In the local achtbein coordinate, the connection, Riemann tensor, Ricci tensor, scalar curvature, and Einstein equations are given by:\footnote{We follow the notations and conventions of \cite{blau2011lecture}. }
\be
\begin{aligned}
	&\Gamma_{lmn} = \frac{1}{2}(d_{lmn} - d_{lnm} + d_{mnl} - d_{mln} + d_{nml} - d_{nlm}),\quad d_{lmn} \equiv e_{l\mu} e_n^\nu \partial_\nu e_m^\mu\\
	&R_{klmn} = 2\partial_{[k} \Gamma_{|mn|l]} + 2\Gamma^a_{\ m[l}\Gamma_{|an|k]} + 2\Gamma^a_{\ [kl]}\Gamma_{mna},\quad R_{km} = R^{\ \, l}_{k\ ml},\qquad R =  R^k_{\ k}\\
	&R_{km} - \frac{1}{2}R\delta_{km} + \Lambda \delta_{km} = 0,\quad k,l,... = 1,2,3,...,8. \\
\end{aligned}
\label{TetradFormalism}
\ee
By extracting independent parts of Einstein equations, we get the following equations of motion for the first ansatz:
\be \small 
\begin{aligned}
	&	-16 a^3 b a' b'-8 a^2 b^2 a'^2-4 a^4 b'^2+28 a^4 b^2+8 a^2 b^2+a^4-2 b^4 = 0,&\\
	&	-4 a b^2 a''-12 a b a' b'-4 b^2 a'^2-4 a^2 b b''-4 a^2 b'^2+28 a^2 b^2+a^2+4 b^2 = 0,&\\
	&	16 a^3 b^2 a''+32 a^3 b a' b'+24 a^2 b^2 a'^2+8 a^4 b b''+4 a^4 b'^2-84 a^4 b^2-24 a^2 b^2-a^4+10 b^4 = 0.&\\
\end{aligned}
\label{EinsteinEquations3}
\ee
Only two of them are independent because of Bianchi identity so they are not over-constrained. An analytical study is desirable, but not yet available unfortunately for now, and partial attempts were done in \cite{Bizon:2007zf} for a similar problem with vanishing cosmological constant. In this paper, we will solve them perturbatively at both large-$r$ and small-$r$ regimes, followed by numerical evaluation, in the same manner as \cite{Bobev:2016sap}. By numerical simulation, we are able to find relations between parameters in large-$r$ series and small-$r$ ones.

Remarkably, the equations of motion (\ref{EOMAnsatz1}) for the second metric ansatz are identical to the ones above when imposing $f_3 = f_4 = f_5$. Thus the analysis below applies identically to the two metrics. In a sense, the second metric is a generalization of the first one which we will focus on in the main text.

\subsubsection{Large radius expansion}
As a space with negative cosmological constant, it's natural to perform Fefferman-Graham expansion at large-$r$:
\be
\left\{
\begin{aligned}
	a(r) = e^r A_0 + A_1 + e^{-r}A_2 + \cdots\\
	b(r) = e^r B_0 + B_1 + e^{-r}B_2 + \cdots\\
\end{aligned}
\right. .
\label{UVExpansionAnsatz2}
\ee
Solving the equations of motion order by order, we find that a general solution can be determined by three parameters, which we choose to be $A_0, B_0, $ and $A_7$. The coefficient $A_7$ is dual to the vacuum expectation value (vev) of a corresponding operator on the boundary \cite{Bianchi:2001kw}, which is the stress tensor in our case and thus should vanish. The first several terms are:
\be
\begin{aligned}
	a(r) = A_0 e^r & + e^{-r} \left(\frac{B_0^2}{8 A_0^3}+\frac{A_0}{80 B_0^2}-\frac{1}{5 A_0}\right) \\
	& +\frac{e^{-3 r} \left(-2 A_0^6 B_0^2-39 A_0^4 B_0^4+140 A_0^2 B_0^6+A_0^8-100 B_0^8\right)}{1600A_0^7 B_0^4} +O(e^{-5r}),\\
	b(r)=B_0 e^r & + \frac{e^{-r} \left(-\frac{10 B_0^4}{A_0^4}+\frac{8 B_0^2}{A_0^2}-3\right)}{80 B_0}\\
	& -\frac{e^{-3 r} \left(-2 A_0^6 B_0^2-39 A_0^4 B_0^4+140 A_0^2 B_0^6+A_0^8-100 B_0^8\right)}{1200A_0^8 B_0^3} +O(e^{-5r}). \\
\end{aligned}
\ee
As a special analytical solution, for the round ${\rm AdS}_8$ with spherical foliation, we have $a(r) = b(r) = \sinh r = \frac{1}{2}e^r - \frac{1}{2}e^{-r}$, which corresponds to $A_0 = B_0 = \frac{1}{2},A_1 = B_1 = -\frac{1}{2}$, and all other coefficients vanish.

\subsubsection{Small radius expansion and numerics - NUT}
\begin{figure}[ht]
	\centering
	\includegraphics[width=11cm]{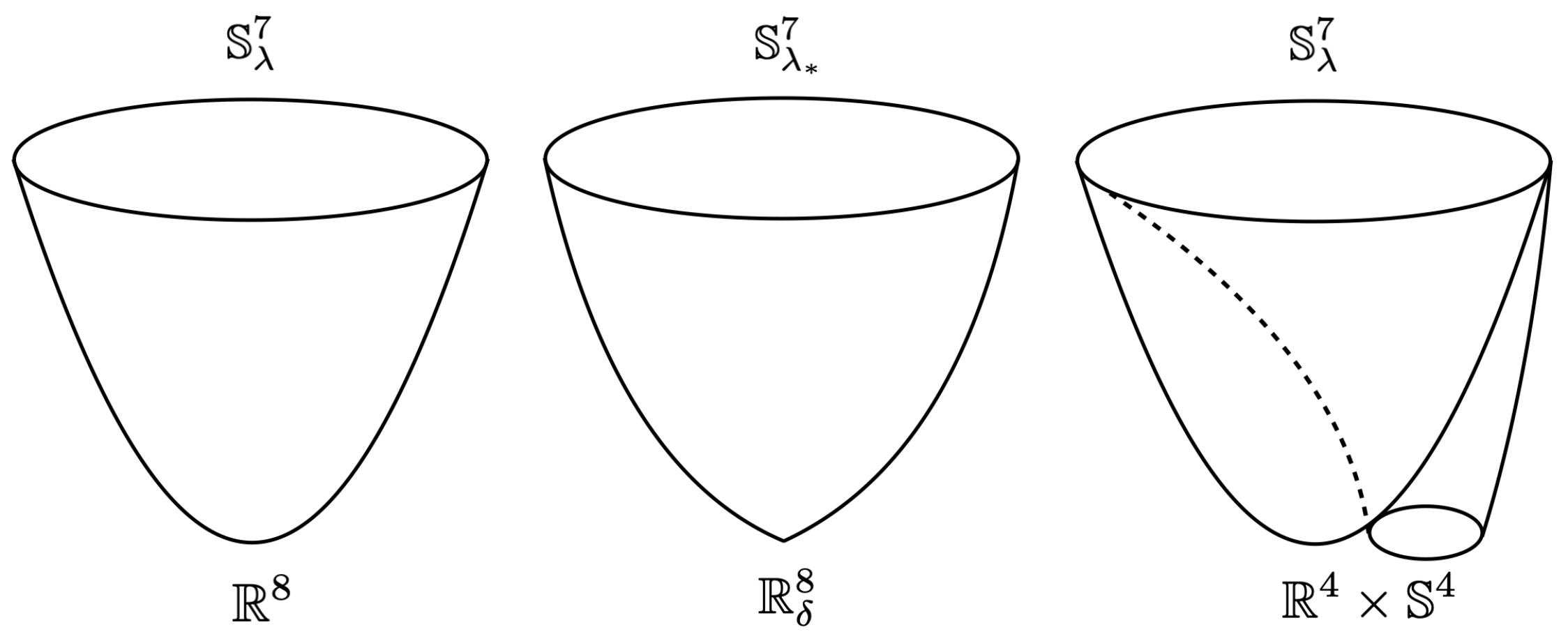}
	\caption{\rm Possible geometries in the bulk. The first one to the left is ``NUT'' geometry, which admits a squashed seven sphere at infinity and caps off smoothly at $r = r_0$. The one on the right is ``Bolt" geometry, which admits a squashed sphere at infinity and caps off smoothly at $r = r_0$, where the geometry is $\mathbb{R}^4\times \mathbb{S}^4$. The one in the middle is a singular ``NUT'' solution which only exists for $\lambda = \lambda_* = \frac{1}{\sqrt{5}}$, it has a conifold singularity at $r = r_0$.}
	\label{Geometries}
\end{figure}

In the small-$r$ region, as $r$ decreases, both $a(r)$ and $b(r)$ decrease, and one of them will hit $0$ at some $r = r_0$, forming a black-hole horizon. Depending on how they hit 0 at $r = r_0$, there're two families of solutions, dubbed by Hawking, ``NUT'' and ``Bolt''. ``NUT'' applies to situations where $a(r)$ and $b(r)$ hits 0 at the same $r_0$, thus the near-horizon geometry looks locally like $\mathbb{R}^8$. While in ``Bolt'', $b(r)$ hits 0 at $r=r_0$ while $a(r_0)$ is finite, for which the near-horizon geometry looks like $\mathbb{R}^4\times \mathbb{S}^4$.  \footnote{On why there's no solution where $a(r)$ hits 0 first when $b(r_0)$ is finite: by expanding at small $r$ and solving equations of motion order by order, one can immediately prove the ansatz doesn't admit a solution. And on why there's no wormhole solution: we've checked it numerically using different initial conditions but failed to find a wormhole. } Besides the normal ``NUT'' and ``Bolt'' geometries, which cap off smoothly at $r = r_0$, there exist a singular ``NUT'' solution which has a conifold singularity at $r = r_0$. The geometries of the solutions are illustrated in Fig.\ref{Geometries}. The two families of solutions are analogous as in $d = 4$ AdS-Taub-NUT/Bolt solution investigated in \cite{Bobev:2016sap}.

For ``NUT'', we assume the following small-$r$ expansion:
\be
\left\{ 
\begin{aligned}
	& a(r) = a_1(r-r_0)^1 + a_2(r-r_0)^2 + ... &\\
	& b(r) = b_1(r-r_0)^1 + b_2(r-r_0)^2 + ... &\\
\end{aligned}
\right. .
\ee
The equations at leading order have two solutions:
\be
\left\{
\begin{aligned}
	&a_1 = \frac{1}{2},&\\
	&b_1 = \frac{1}{2},&
\end{aligned}
\right.\qquad {\rm or}\qquad
\left\{
\begin{aligned}
	&a_1 = \frac{3\sqrt{5}}{10},&\\
	&b_1 = \frac{3}{10}.&
\end{aligned}
\right.
\label{twoChoicesAnsatz2}
\ee
For the first choice, we solved the equations up to $O(r-r_0)^{13}$, and found the series has only one free parameter, which is $a_3$, and the first several orders are
\be \small 
\begin{aligned}
	a(r) = &\frac{\rho}{2}+a_3 \rho^3+\frac{\left(-14832 a_3^2+1932 a_3-49\right) \rho^5}{2160}+\frac{1}{816480}\left(44570304 a_3^3-7822224 a_3^2\right.\\
	&\left.+434448 a_3-7595\right)
	\rho^7 + O\left(\rho^9\right), \\
	b(r) = &\frac{\rho}{2}+\left(\frac{7}{36}-\frac{4 a_3}{3}\right) \rho^3+\frac{\left(71424 a_3^2-9744 a_3+343\right) \rho^5}{6480}+\frac{1}{816480}\left(-83054592 a_3^3\right.\\
	&\left.+15344640
	a_3^2-906192 a_3+17101\right) \rho^7+O\left(\rho^9\right), \\
\end{aligned}	
\label{SmallrExpansionNUT}
\ee
where we used the short-handed notation $\rho \equiv r-r_0$. Note that when choosing $a_3 = \frac{1}{12}$, one obtains the round ${\rm AdS}_8$ solution:
\be
a(r) = b(r) = \sinh \rho\quad \Rightarrow \quad ds^2 = d\rho^2 + \sinh^2\rho\, ds^2_{\mathbb{S}^7}.
\label{RoundSphere}
\ee 
For the second choice in (\ref{twoChoicesAnsatz2}), up to order $O(r-r_0)^{13}$, the solution is fixed, with no free parameter, whose leading order expansions can be identified with hyperbolic sine functions:
\be
\begin{aligned}
	&a(r) = \frac{3\sqrt{5}}{10}\left(\rho +\frac{\rho ^3}{6}+\frac{\rho ^5}{120}+\frac{\rho ^7}{5040}+\frac{\rho^9}{362880} + O\left(\rho ^{11}\right)\right) \rightarrow  \frac{3\sqrt{5}}{10}\sinh\rho, &\\
	&b(r)  = \frac{3}{10}\left(\rho +\frac{\rho ^3}{6}+\frac{\rho ^5}{120}+\frac{\rho ^7}{5040}+\frac{\rho ^9}{362880}+O\left(\rho^{11}\right)\right)\rightarrow \frac{3}{10}\sinh\rho. &\\	
\end{aligned}
\label{branch1ansatz2IR}
\ee
As can be checked, the asymptotic boundary of the above solution is the Einstein squashed sphere (\ref{SquashedSphereMetric2}) with $\lambda^2_* = 1/5$. This geometry is special, in one sense, it's the only singular solution that integrates to infinity, which can be shown by its diverging Kretschmann scalar at $r \rightarrow r_0$:
\be 
R = -56,\quad  R_{\mu\nu}R^{\mu\nu} = 392,\quad R_{\mu\nu\rho\sigma}R^{\mu\nu\rho\sigma} = 112 + \frac{2^{12}}{3^3\sinh^4\rho}.
\label{Curvatures}
\ee 
If we zoom in to the near-horizon limit, where $a(r) \approx \frac{3\sqrt{5}}{10}\rho$ and $b(r)\approx  \frac{3}{10}\rho$, we can evaluate:
\be 
R = 0,\quad  R_{\mu\nu}R^{\mu\nu} = 0,\quad R_{\mu\nu\rho\sigma}R^{\mu\nu\rho\sigma} = \frac{2^{12}}{3^3\rho^4 }.
\ee 
The appearance of a singular solution is not typical in Taub-NUT solution, since the introduction of nut charge actually alleviates the singular behavior in a Schwarzschild black hole. A similar singular solution is found in \cite{Aharony:2019vgs} when the boundary dimension satisfies $d \ge 9$.

Given the function values of $a(r),b(r)$ and their first-order derivatives at $\rho = 0$, we can solve the differential equations (\ref{EinsteinEquations3}) numerically. In general, for a randomly-chosen parameter $a_3$, it's not guaranteed that both $a(r)$ and $b(r)$ can be integrated up to infinity without hitting 0, otherwise one has a compact spindle-like space instead of an ${\rm AlAdS}_8$ one. In this paper, we focus on the solutions where neither $a(r)$ nor $b(r)$ has zeros, a bulk solution with asymptotic squashed sphere boundaries as (\ref{SquashedSphereMetric2}), where the squashing parameter can be identified with the UV expansion coefficients in (\ref{UVExpansionAnsatz2}) as
\be
\lambda = \lim_{r\rightarrow\infty} \frac{b(r)}{a(r)} = \frac{B_0}{A_0} .
\ee
\begin{figure}[!t]
	\centering	\includegraphics[width=7cm]{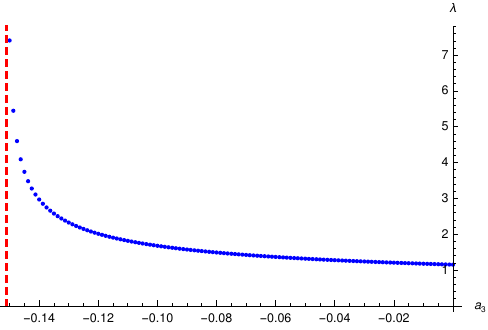}
	\qquad \qquad 
	\includegraphics[width=7cm]{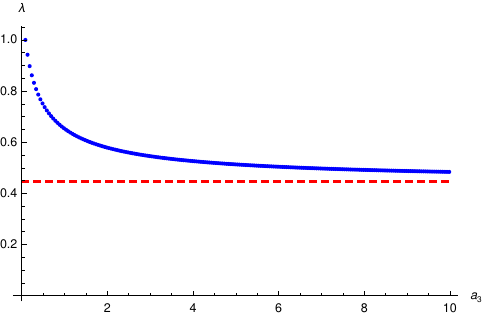}
	\hfil
	\includegraphics[width=7cm]{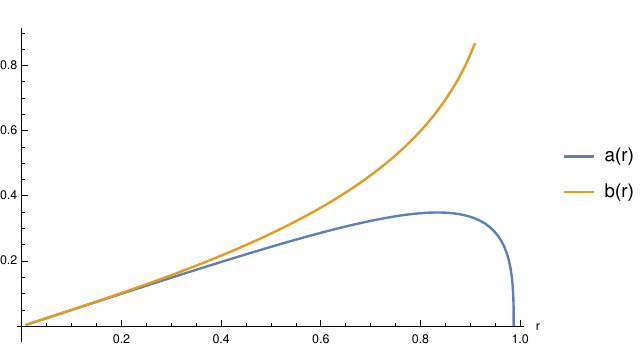}
	\qquad \qquad 
	\includegraphics[width=7cm]{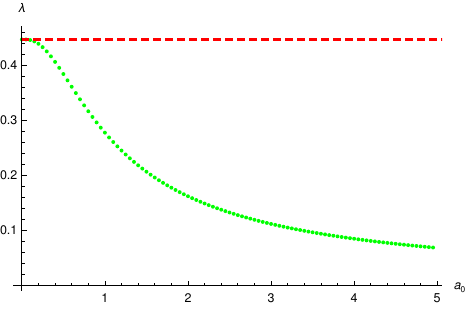}
	\caption{{\rm Upper: the relation between $\lambda$ on the boundary and $a_3$ at small-$r$, the verticle dashed line is $a_3 = a_3^*$, and the horizontal dashed line is $\lambda=\lambda_*=1/\sqrt{5}$. Downleft: numerical solution for NUT with initial value $a_3 = -0.5$. Downright: relation between $\lambda$ and $a_0$ for Bolt, the dashed line is $\lambda = \lambda_*$.}}
	\label{UVIRRelationAnsatz2}
\end{figure}
We start from $a_3 = \frac{1}{12}$ and integrate numerically to infinity, we get $\lambda = 1$, as expected. With different $a_3$, this procedure establishes a one-to-one correspondence between the squashing parameter $\lambda$ at asymptotic boundary and the initial condition $a_3$ at small-$r$, as shown in Fig.\ref{UVIRRelationAnsatz2}. Increasing the value of $a_3$ from $1/12$, the corresponding $\lambda$ decreases monotonically from $1$ to $\lambda_* = 1/\sqrt{5}\approx 0.447$; we then decrease $a_3$, and $\lambda$ increases monotonically from 1 to $\infty$ and finally stops at a special value $a_3^*\approx -0.151394$. For any $a_3<a_3^*$, $a(r)$ doesn't integrate to infinity, thus is of no interest to us. An example for $a_3 = -0.5 < a_3^*$ is shown in Fig.\ref{UVIRRelationAnsatz2}.

The NUT solution also has interesting implications in terms of holographic RG flow. It interpolates an asymptotically locally squashed AdS$_8$ at the UV region and a flat $\mathbb{R}^8$ in the IR region. The interesting fact is that from a UV theory with different squashing parameters $\lambda$, the RG flow converges to the same IR theory, which, from the leading order coefficients in (\ref{SmallrExpansionNUT}), lives in $\mathbb{R}^8$. The picture here is similar to the holographic uniformization discovered in \cite{Anderson:2011cz}, which is worth further understanding.

\subsubsection{Small radius expansion and numerics - Bolt}
The geometry of ``Bolt'' solution is supposed to be $\mathbb{R}^4\times \mathbb{S}^4$ at small-$r$ limit, where $a(r_0)>0$ is the finite radius of $\mathbb{S}^4$, and $b(r)$ goes to zero smoothly, forming $\mathbb{R}^4$ together with the radial coordinate. The geometry indicates the following small-$r$ expansion:
\be
\begin{aligned}
	& a(r) = a_0 + a_1(r-r_0)^1 + a_2(r-r_0)^2 + ... &\\
	& b(r) = b_1(r-r_0)^1 + b_2(r-r_0)^2 + ... &\\
\end{aligned} .
\ee
Repeating the procedure as above, the solutions can be obtained in any order, determined by one free parameter $a_0>0$. \footnote{Taking the limit $a_0 \rightarrow 0$, the geometry becomes closer to NUT solutions, as can be seen from the corresponding values of $\lambda$.} The leading terms are as follows

\be \small 
\begin{aligned}
	a(r) = & a_0-\frac{\left(-7 a_0^2-3\right) \rho^2}{8 a_0}-\frac{\left(49 a_0^4+98 a_0^2+39\right) \rho^4}{384 a_0^3}\\
	&\quad  -\frac{\left(-7889 a_0^6-15141
		a_0^4-10241 a_0^2-2379\right) \rho^6}{46080 a_0^5}+O\left(\rho^8\right),\\
	b(r) = & \frac{\rho}{2}-\frac{\rho^3}{12 a_0^2}-\frac{\left(-49 a_0^4-70 a_0^2-26\right) \rho^5}{480 a_0^4}\\
	&\quad  -\frac{\left(10290 a_0^6+20531 a_0^4+14000
		a_0^2+3224\right) \rho^7}{80640 a_0^6}+O\left(\rho^9\right).\\
\end{aligned}
\ee
The functions can be integrated out to infinity for any $a_0 > 0$, and the map between $a_0$ and $\lambda$ is shown in Fig.\ref{UVIRRelationAnsatz2}. As shown in the plot, as $a_0$ increases from 0 to $\infty$, $\lambda$ decreases from $\lambda_*$ to $0$ monotonically. Combine the $\lambda\sim a_0$ plot for ``Bolt'' solutions with $\lambda \sim a_3$ plot for ``NUT'' solutions, we find that for a squashed seven-sphere metric (\ref{SquashedSphereMetric2}) with any $\lambda>0$, there exists one unique bulk solution whose asymptotic boundary is squashed seven-sphere with $\lambda$. For $0<\lambda<\lambda_*$, the bulk is ``Bolt''; for $\lambda\ge \lambda_*$, the bulk is ``NUT''. The two analytical solutions, which have $\lambda = 1$ and $\lambda = \lambda_*=1/\sqrt{5}$, are both ``NUT'' spaces. Inversely speaking, the bulk metric is uniquely determined by $\lambda$, so there will not be an analog of Hawking-Page transition in our geometry. But it's still meaningful to evaluate the bulk free energy $F_{\rm  bulk}(\lambda)$ as a function of the squashing parameter, which we will do in the next section.

\subsection{Free energy}
\begin{figure}[t]
	\centering
	\includegraphics[width=7cm]{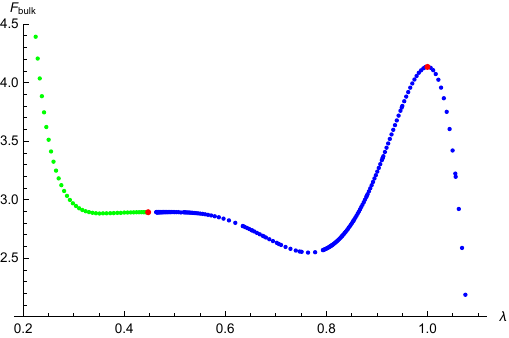}
	\caption{{\rm Numerical renormalized free energy $F_{\rm bulk}(\lambda)$ for NUT and Bolt spaces. The values of $F_{\rm bulk}$ are divided by $\frac{\ell^6}{G_N}$. Where blue points show NUT solutions, green points are Bolt solutions, and red points are the two special values $\lambda = 1, \lambda_*$ given in \eqref{AnalyticalBulkFE}. }}
	\label{FscLambda}
\end{figure}

In the semi-classical limit, the partition function localizes to solutions of Einstein equations, then the Euclidean gravitational free energy is reduced to the on-shell action:
\be 
F_{\rm bulk} = - \log \int \cD g_{\mu\nu} e^{-S_E[g_{\mu\nu}]} \xlongequal[\rm approximation]{\rm classical} - \log e^{-S_E[g_{\mu\nu}]_{\rm on-shell}} = S_E[g_{\mu\nu}]_{\rm on-shell}.
\ee 
We focus on Euclidean gravitational field without matter in the bulk, whose free energy with the GHY term on the boundary is given by \cite{Gibbons:1976ue}:
\be
S_{\rm EH} + S_{\rm GHY}= -\frac{ 1 }{16\pi G_N}\int_{\cM_8} d^{8}x\sqrt{g}\left(R -2\Lambda\right) - \frac{ 1 }{8\pi G_N}\int_{\partial\cM_8} d^7 x\sqrt{g^{(7)}}K,
\ee
where $g^{(7)}_{ab}$ is the induced metric on $\partial\cM_8$, which we choose to be a surface with constant large $r$, and $K$ is the trace of extrinsic curvature tensor on the boundary:\footnote{In our convention, Greek letters starting from $\mu$ are bulk indices, and Latin letter starting from $a$ are boundary indices. }\cite{blau2011lecture} 
\be 
g_{\mu\nu}^{(7)} \equiv g_{\mu\nu} - n_\mu n_\nu ,\quad K_{\mu\nu} \equiv (g^{(7)})_\mu^\rho (g^{(7)})_\nu^\sigma \nabla_\rho n_\sigma.
\ee 
Here $n^\mu$ is the unit normal vector of the boundary. By plugging in the metric ansatz, we obtain the action and its boundary term:
\be\begin{aligned}
	S_{\rm EH} &= \int_0^{R_{\partial}} dr\ \frac{4\pi^3 \ell^5 b}{3G_N}\left( 48 a^3 b a' b'+24 a^2 b^2 a'^2+16 a^3 b^2 a''+12 a^4 b'^2+12 a^4 b b''\right. \\
	&\left. \qquad\qquad \qquad  -84 a^4
	b^2-24 a^2 b^2-3 a^4+6 b^4 \right),\\
	S_{\rm GHY} &= - \frac{16\pi^3 \ell^6 a^3b^2}{3G_N} \left( 4 b a'+3 a b' \right).\\
\end{aligned}\label{GravAction} \ee 
For general AlAdS space, the free energy above is proportional to the volume of $\cM_8$, which is infinite, corresponding to a UV divergence. So we need to regularize the divergence by holographic renormalization. \cite{Henningson:1998gx, Henningson:1998ey, Balasubramanian:1999re} The counter-terms for $d\le 7$ are given by \cite{Clarkson:2002uj, Astefanesei:2004kn, Emparan:1999pm}\footnote{See \cite{Bueno:2022log} for a recent algorithm for deducing the counter terms up to arbitrarily large order.}
\be \small 
\begin{aligned}
	S_{\rm ct} =&  \frac{ 1 }{8\pi G_N} \int_{\partial \cM} d^d x\sqrt{g^{(d)}}\left[(d-1) + \frac{1}{2(d-2)}\cR + \frac{1}{2(d-4)(d-2)^2}\left(\cR_{ab}\cR^{ab}-\frac{d}{4(d-1)}\cR^2\right)\right.\\
	& \left.	+\frac{1}{(d-2)^3(d-4)(d-6)}\left(\frac{3d+2}{4(d-1)}\cR \cR_{ab}\cR^{ab}-\frac{d(d+2)}{16(d-1)^2}\cR^3 - 2\cR^{ab}\cR^{cd}\cR_{acbd}\right.\right.\\
	&\left.\left.-\frac{d}{4(d-1)}\nabla_a\cR\nabla^a \cR +\nabla^c\cR^{ab}\nabla_c\cR_{ab}\right)\right],
\end{aligned}
\ee
where $\cR_{abcd}, \cR_{ab}, \cR$ are the Riemann tensor, Ricci tensor, and Ricci scalar of the boundary. Plugging in the metric ansatz, we get:
\be \small  \begin{aligned}
	S_{\rm ct} & = \frac{\pi^3 \ell^6 }{12000a^8b^3G_N} \left( 384000 a^{12} b^6+9600 a^{12} b^4+40 a^{12} b^2+76800 a^{10} b^6-4480 a^{10} b^4 \right. \\
	&\left.  +40 a^{10}
	b^2-19200 a^8 b^8-160 a^8 b^6+2774 a^8 b^4-6400 a^6 b^8+1504 a^6 b^6+4000 a^4
	b^{10} \right. \\
	&\left.  -13900 a^4 b^8+23200 a^2 b^{10}+7 a^{12}-11000 b^{12} \right).
\end{aligned}\ee
In the following evaluations, we set the Newton constant to be unit. Recall that there're two special values of $\lambda^2 = 1, 1/5$ for which the metric $g_{\mu\nu}$ is known analytically (\ref{RoundSphere}, \ref{branch1ansatz2IR}), the free energies can also be calculated analytically by $F_{\rm bulk}(\lambda) = S_{\rm EH} + S_{\rm GHY} + S_{\rm ct}$, which are:
\be
F_{\rm bulk}(1) = \frac{2\pi^3}{15} \frac{\ell^6}{G_N} \approx 4.134  \frac{\ell^6}{G_N}  ,\qquad F_{\rm bulk}\left(\frac{1}{\sqrt{5}}\right) = \frac{2\cdot 3^6 \pi ^3}{5^6}  \frac{\ell^6}{G_N}  \approx 2.893  \frac{\ell^6}{G_N} .
\label{AnalyticalBulkFE}
\ee
For general $\lambda$, $F_{\rm bulk}(\lambda)$ can be obtained only numerically. By large-$r$ expansion, we can check that the diverging part of $S_{\rm EH} + S_{\rm GHY}$ and $S_{\rm ct}$ exactly cancel with each other, which have the same diverging structure:
\be \small 
\begin{aligned}
	S_{\rm EH} &  + S_{\rm GHY} \sim  -S_{\rm ct} \sim \frac{B_0\pi^3 \ell^6 }{G_N}\left[-32 A_0^4 B_0^2 e^{7 r}+\frac{6}{5} e^{5 r} \left(8 A_0^2 B_0^2+A_0^4-2 B_0^4\right)\right.\\
	& +\frac{1}{600} e^{3 r} \left(\frac{100
		B_0^6}{A_0^4}-\frac{160 B_0^4}{A_0^2}+\frac{A_0^4}{B_0^2}-112 A_0^2-4 B_0^2\right)\\
	& \left.	+\frac{e^r }{9600}\left(\frac{11000
		B_0^8}{A_0^8}-\frac{23200 B_0^6}{A_0^6}+\frac{13900 B_0^4}{A_0^4}-\frac{1504 B_0^2}{A_0^2}-\frac{40
		A_0^2}{B_0^2}-\frac{7 A_0^4}{B_0^4}-2774\right)\right].
\end{aligned}
\label{IRDivergenceExpansion}
\ee
We use the notation ``$\sim$'' to denote that the equation holds up to vanishing terms at large $r$. The numerical renormalized free energy is shown in Fig.\ref{FscLambda}. As $\lambda\rightarrow 0$ or $\lambda\rightarrow +\infty$, $F_{\rm bulk}$ diverges as expected. At $\lambda = 1$, there is a local maximum, consistent with the holographic F-theorem. Up to the precision of our numerics, near $\lambda_* = \frac{1}{\sqrt{5}}$, the free energy varies quite slowly, but we are not able to determine the critical behaviors explicitly.

Since we're integrating numerically starting from $r = 0$, the numerical error keeps accumulating and explodes at around $r \approx 5$, thus one has to stop integrating somewhere at $r = R < 5$, then the action looks like:
\be 
S_{\rm EH} + S_{\rm GHY} = A e^{7R} + B e^{5R} + C e^{3R} + D e^{R} + E + O\left( e^{-R}\right).
\ee 
The diverging terms are canceled by the counter term, and what we're interested in is the finite term $E$. To extract $E$, we take a set of cutoff values $R$ and numerically fit it with the function of the form $E + F e^{-R} + G e^{-3R}+...$, until the discrepancy from the analytical results (\ref{AnalyticalBulkFE}) is small enough.

In the end, we would like to mention that the gravitational action and boundary term of the second metric ansatz with $f_3=f_4=f_5$ are identical to those posted above (\ref{GravAction}), thus the renormalized action is identical to the one we posted in Fig.\ref{FscLambda}.

\section{Field theory story}
\label{BoundaryTheory}

It's also interesting to study the free energies of quantum field theories. To look for the correct holographic correspondence of the Einstein gravity theory, one has to go to string or M-theory and identify the boundary CFT, which is typically a strongly-coupled CFT without supersymmetry. Instead, we will study the universal properties of the free energies dictated by conformal symmetry. Directly making use of the conformal symmetry, we can study universal properties shared by all CFTs living on the squashed sphere as in \cite{Bobev:2017asb}. Some of the universal properties can also be found by studying some toy models as in \cite{Hartnoll:2005yc, Anninos:2012ft, Anninos:2013rza, Bobev:2016sap, Bobev:2017asb}.

The main quantity that we are interested in is the free energy $F_{\rm CFT}(\lambda)$ of CFTs as a function of the squashing parameter that the field theory lives. In the first part of this section, we calculate $F_{\rm CFT}''(\lambda)\Big|_{\lambda = 1}$ for general CFT living on the squashed seven-sphere of the first kind (\ref{SquashedSphereMetric2}), which can be evaluated by integrating the two-point function of stress tensors on the sphere. This property is totally determined by the conformal symmetry and thus independent of the details of the theories such as couplings.

To support the study of the correspondence both around and away from $\lambda = 1$, we study in addition two free CFTs, the conformally coupled scalar, and the free fermion respectively. Using these toy models, we can evaluate $F_{\rm CFT}'''(\lambda)\Big|_{\lambda = 1}$ both analytically and numerically, which is too complicated to be evaluated as an integrated three-point function of stress tensors. Besides, we can also study the strongly deformed scaling of the field theories numerically, which can also be compared with the bulk.

Let's clarify our results for the two metric ansatzes. The result in the first subsection only works for the first ansatz which is easier to deal with, we leave a detailed study of the second ansatz for future work. But we do expect they are the same because our results for the toy models are identical for them both due to the coincidence of spectrum on the two squashed sphere metrics. What's more, the results show a very good coincidence with the bulk free energy, which is also identical for the two metrics.

\subsection{Partition function on deformed manifold}

The partition function of a general CFT on the boundary of manifold $\cM_{d+1}$ with Euclidean boundary metric $g_{ab}$ is given by:\footnote{In this section, we follow the notation of \cite{Bobev:2017asb} for the free energy. Note that the quantity $\widetilde{F}$ in $F$-theorem \cite{Pufu:2016zxm} in \eqref{FreeEnergySd} is different from the free energy by a sign, which happens to be $+1$ for $d = 7$ of our interest. } 
\be
Z = \int \cD \varphi\; e^{-S[\varphi,g_{ab}]},\quad F \equiv -\ln Z.
\ee
We can couple the field theory to the gravitational background by squashing the metric. As is mentioned in the introduction, squashing background metric is equivalent to adding a marginal deformation (\ref{Deformation}) generated by the stress tensor, where we parametrize the squashing by a parameter $\epsilon$ as follows:
\be 
g_{ab} = g_{ab}^{(0)} + \epsilon h_{ab},\quad g^{ab} = (g^{(0)})^{ab} + \epsilon h^{ab} + O(\epsilon^2) \quad \Rightarrow \quad  h^{ab} = -h_{cd}(g^{(0)})^{ac}(g^{(0)})^{bd}.
\ee 
The coupling of field theory with background metric, or equivalently the effect of the induced RG flow, is encoded in correlation functions of the stress tensor. To start with, the first-order derivative of the free energy is given by integrating the one-point function of the stress tensor: \cite{Bobev:2017asb}
\be 
F'(\epsilon)\Big|_{\epsilon = 0} = - \frac{1}{2} \int d^dx \sqrt{g^{(0)}} h^{ab}(x) \langle T_{ab}(x)\rangle_{\partial \cM}.
\ee  
In odd-dimensional CFTs, out setup included, one point function of stress tensor vanishes as is required by conformal symmetry, thus $F'(0) = 0$. The second derivative of free energy $F_{\rm CFT}(\epsilon)$ is given by the integrated two-point function of stress tensor:
\be
F''(\epsilon)\Big|_{\epsilon=0} = -\frac{1}{4}\int d^dx d^dy\sqrt{g^{(0)}(x)g^{(0)}(y)} h^{ab}(x)h^{cd}(y)\langle T_{ab}(x)T_{cd}(y)\rangle_{\partial \cM}.
\label{Integrated2ptFunc}
\ee
The integrand is evaluated on the boundary $\partial \cM$ with deformations turned off. To evaluate the two-point function on $\partial \cM$, we take a conformal map $f:\partial\cM\rightarrow \mathbb{R}^d$ which relates the line element as $$f^*(ds^2_{\mathbb{R}^d} ) = \Omega^2(x)ds^2_{\partial\cM},$$ where $f^*$ is the pullback and $\Omega^2(x)$ is the corresponding conformal factor. In odd dimensions, the stress tensors are transformed under the conformal map through \cite{Simmons-Duffin:2016gjk}
\be \label{defMab}
T_{ab}(x) = \Omega^{d-2}M^{\bar{a}\bar{b}}_{ab}T_{\bar{a}\bar{b}}(X),\quad {\rm where} \ M^{\bar{a}\bar{b}}_{ab}\equiv \frac{\partial X^{\bar{a}}}{\partial x^a}\frac{\partial X^{\bar{b}}}{\partial x^b},\quad X^{\bar{a}} \in \mathbb{R}^d,\ x^a \in \partial \cM.
\ee
Thus the stress tensor two-point function on $\mathbb{S}^7$ is related to that on $\mathbb{R}^7$ by
\be
\langle T_{ab}(x)T_{cd}(y)\rangle_{\mathbb{S}^7} =                          \Omega^5(x)\Omega^5(y)M_{ab}^{\bar{a}\bar{b}}M_{cd}^{\bar{c}\bar{d}}\langle T_{\bar{a}\bar{b}}(X)T_{\bar{c}\bar{d}}(Y)\rangle_{\mathbb{R}^7}.
\label{StressTensor2ptFunction}
\ee
And the two-point function on $\mathbb{R}^d$ is well-known in the literature by using conformal symmetry: \cite{Osborn:1993cr,Erdmenger:1996yc}
\be
\begin{aligned}
	& \langle T_{\bar{a}\bar{b}}(X)T_{\bar{c}\bar{d}}(Y)\rangle_{\mathbb{R}^d}     = C_T \frac{I_{\bar{a}\bar{b}\bar{c}\bar{d}}(X-Y)}{|X-Y|^{2d}},\quad I_{\bar{a}\bar{b}, \bar{c}\bar{d}}(X) =\cE_{\bar{e}\bar{f},\bar{c}\bar{d}}I_{\bar{a}\bar{e}}I_{\bar{b}\bar{f}}, \\
	& \cE_{\bar{a}\bar{b},\bar{c}\bar{d}} = \frac{1}{2}(\delta_{\bar{a}\bar{c}}\delta_{\bar{b}\bar{d}} + \delta_{\bar{a}\bar{d}}\delta_{\bar{b}\bar{c}}) - \frac{1}{d}\delta_{\bar{a}\bar{b}}\delta_{\bar{c}\bar{d}},\quad I_{\bar{a}\bar{b}}(X) = \delta_{\bar{a}\bar{b}} - 2 \frac{X_{\bar{a}}X_{\bar{b}}}{X^2}.\\
\end{aligned}
\label{TTCorrelation}
\ee
There exists a natural conformal map between $\mathbb{S}^d$ and $\mathbb{R}^d$: the stereographic projection, for which the conformal factor is given by:
\be \Omega = \frac{1}{2}\left( 1 + X_{\bar{a}} X^{\bar{a}} \right), 
\label{ConformalFactorOmega}
\ee 
thus one can always evaluate the reaction of the free energy under metric deformation. But for a specific metric on the sphere, the detailed construction of the global map may be different, more discussions and explicit constructions of the conformal map can be found in Appendix \ref{ConformalMapping}. Using the conformal map, one can work out the matrix $M_{ab}^{\bar{a}\bar{b}}$ defined in \eqref{defMab}, which facilitates the integral. In this section, we will investigate two different squashings, one is the SU(2) bundled seven spheres which is the theme of this paper, and another one is the U(1) bundled squashed $\mathbb{S}^{2d+1}$ which is also of great interest.

\subsubsection{SU(2) bundle squashed $\mathbb{S}^7$}    

To apply the formalism above to our metric (\ref{SquashedSphereMetric2}) on $\mathbb{S}^7$, we can relate the two squashing parameters as follows: $$\epsilon \equiv \lambda^2 -1,$$ for which one finds the only non-vanishing components of the perturbative inverse metric $h^{ab}$ are:
\be
h^{\theta\theta} = -4,\quad h^{\phi\phi} = h^{\psi\psi} = -\frac{4}{\sin^2\theta},\quad h^{\phi\psi} = h^{\psi\phi} = \frac{4}{\tan\theta \sin\theta}.
\label{hComp}
\ee 
Now we know everything in our integral \eqref{Integrated2ptFunc}, which is the double integral on $\mathbb{S}^7$. To evaluate it, one doesn't need to perform a 14-dimensional integration, instead, one fixes one stress tensor on the south pole and integrates the other stress tensor over the sphere. \cite{Pufu:2016zxm, Cardy:1988cwa, Bobev:2017asb} What considered in the references is the correlation function of scalar operators, thus the trick is easily justified because the scalar perturbation preserves the symmetry of the sphere. However, we're dealing with the correlation function of stress tensors, whose deformation coefficients are tensors with coordinate-dependent components as shown in \eqref{hComp}, so it is not understood why the trick still works. Luckily it works well and corresponds to the other calculations that we did in this paper. According to the trick, we put one point $y^a$ on the south pole, and only integrate on $x^a$:
\begin{equation*}
    \begin{aligned}
    F''(\epsilon)\Big|_{\epsilon=0} &= - \frac{C_T}{4}V_{\mathbb{S}^7} \int d^7x \sqrt{g^{(0)}(x)} h^{ab}(x)h^{cd}(0) \langle T_{ab}(x)T_{cd}(0)\rangle_{\mathbb{S}^7}\\
    &  = - \frac{C_T}{4}V_{\mathbb{S}^7} h^{cd}(0)\Omega^{5}(0)M_{cd}^{\bar{c}\bar{d}}(0)\int d^7x \sqrt{g^{(0)}(x)} \left[h^{ab}(x)\Omega^{5}(x)M_{ab}^{\bar{a}\bar{b}}(x)\frac{\cI_{\bar{a}\bar{b},\bar{c}\bar{d}}(X)}{|X|^{14}}\right],
\end{aligned}
\end{equation*}
where $V_{\mathbb{S}^7} = \frac{\pi^4}{3}$ is the volume of the unit seven sphere, coming from identifying the integrand with arbitrary $y^\mu \in \mathbb{S}^7$ to that with $y^\mu$ fixed at the south pole. The south pole of the sphere corresponds to the origin of $\mathbb{R}^7$, where the quantities are:
\be 
    \Omega(0) = \frac{1}{2},\quad M^{\bar{5}\bar{5}}(0) = M^{\bar{6}\bar{6}}(0) = M^{\bar{7}\bar{7}}(0) = \frac{1}{4},\quad M^{\bar{c}\bar{d}} \equiv h^{cd}M_{cd}^{\bar{c}\bar{d}}.
\ee 
Then the integrand, which we denote $I$ is given by:
\be 
    I = - \frac{C_T V_{\mathbb{S}^7}}{2^9} \sqrt{g^{(0)}(x)} \frac{(1-x^8)^9}{(x^{\bar{a}} x_{\bar{a}})^7} \sum_{\bar{c} = (5,6,7)} M^{\bar{a}\bar{b}}I_{\bar{c}\bar{c};\bar{a}\bar{b}}(X),\quad x^A \in \mathbb{R}^8, \ \bar{a} = 1,2,\cdots 7.
\ee 
    Since the integrand is rather complicated, in order to evaluate it analytically, we split the contraction into the following three terms:
\be 
    \begin{aligned}
        \sum_{\bar{c} = (5,6,7)} M^{\bar{a}\bar{b}}I_{\bar{c}\bar{c};\bar{a}\bar{b}}(X) &= \left( M^{\bar{5}\bar{5}} + M^{\bar{6}\bar{6}} + M^{\bar{7}\bar{7}} - \frac{3}{7} M^{\bar{a}\bar{a}} \right) - \frac{4(1-x^8)^2}{x^{\bar{a}} x_{\bar{a}}} \sum_{\bar{c} = 1}^7 \sum_{\bar{d} = 1}^3 M^{\bar{c}\bar{d}} X_{\bar{c}}X_{\bar{d}} \\
        & + \frac{4(1-x^8)^2}{(x^{\bar{a}} x_{\bar{a}})^2 } M^{\bar{c}\bar{d}} X_{\bar{c}}X_{\bar{d}}( x^{\bar{5}} x_{\bar{5}} + x^{\bar{6}} x_{\bar{6}} + x^{\bar{7}} x_{\bar{7}} ).
    \end{aligned}
\ee 
Thus we have $I = I_1 + I_2 + I_3$ where
\be\begin{aligned}
    I_1 &= - \frac{C_TV_{\mathbb{S}^7} }{2^9} \sqrt{g^{(0)}(x)} \frac{(1-x^8)^9}{(x^{\bar{a}} x_{\bar{a}})^7} \left( M^{\bar{5}\bar{5}} + M^{\bar{6}\bar{6}} + M^{\bar{7}\bar{7}} - \frac{3}{7} M^{\bar{a}\bar{a}} \right)  \\
    I_2 &= + \frac{C_TV_{\mathbb{S}^7}}{2^7} \sqrt{g^{(0)}(x)} \frac{(1-x^8)^{11}}{(x^{\bar{a}} x_{\bar{a}})^8} \sum_{\bar{c}=1}^7\sum_{\bar{d}=1}^3M^{\bar{c}\bar{d}} X_{\bar{c}}X_{\bar{d}} \\
    I_3 &= - \frac{C_TV_{\mathbb{S}^7}}{2^7} \sqrt{g^{(0)}(x)}  \frac{(1-x^8)^{11}}{(x^{\bar{a}} x_{\bar{a}})^{9}} M^{\bar{c}\bar{d}} X_{\bar{c}}X_{\bar{d}}( x^{\bar{5}} x_{\bar{5}} + x^{\bar{6}} x_{\bar{6}} + x^{\bar{7}} x_{\bar{7}} ). \\
\end{aligned}\label{FullIntegral}\ee
Substituting $h^{ab}$ from \eqref{hComp}, $\Omega^2$ from \eqref{ConformalFactorOmega}, and $M_{ab}^{\bar{a}\bar{b}}$ from the conformal map given by combining \eqref{StereographicProjection} and \eqref{EmbeddingMapS7}, we can express the integrand in terms of $\mathbb{S}^7$ coordinates $(\mu, \Theta, \Phi, \Psi, \theta, \phi, \psi)$. Since the three integrals above are all divergent near the south pole, the order of variables to be integrated out is important. Here we present our procedure, which might be not a unique way to get the correct answer. 

The dependence of the integrands on $(\Theta, \Phi, \Psi)$ is very simple, so we can integrate over them directly. Then we integrate out $\theta$ followed by $\mu$, both of which are convergent, and the resulting function only depends on $\chi \equiv \phi + \psi$. The final function is divergent, but the integration is ignorant of the divergence, which is similar to the example below:
\be 
    \int_{-1}^1 \frac{1}{x^2}dx = -\frac{1}{x}\Big|_{-1}^1 = -2.
\ee 
By going through the procedures above, one can obtain that
\be 
\begin{aligned}
    & F''(\epsilon)\Big|_{\epsilon=0} = \int (I_1 + I_2 + I_3) = - \frac{29\pi^8}{3780}  C_T + \frac{\pi^8}{1890} C_T + 0 =  - \frac{\pi^8}{140}C_T\\
    & \quad \Rightarrow \quad  F''(\lambda)\Big|_{\lambda = 1} = 4F''(\epsilon)\Big|_{\epsilon=0} = - \frac{\pi^8}{35} C_T.\\
\end{aligned}
\label{FppSU2Bundle}
\ee 
In the last line above, we used the relation between the squashings $\epsilon = \lambda^2 - 1$. The result is universal since it's only related to conformal symmetry and thus applies to all CFTs. It must be reproduced in free field theories as well as the holographic theory, as we will see.

\subsubsection{U(1) bundle squashed $\mathbb{S}^{2k+1}$ }    
Spheres as U(1) bundles over complex projective spaces are very interesting cases and have been investigated partially in the literature. \cite{Bobev:2017asb, Bueno:2018yzo, Bueno:2020odt} As in the last section, we can obtain the second-order derivative of free energy in terms of squashing parameter for any odd-dimensional sphere with a U(1) bundle. The special cases for $\mathbb{S}^3$ and $\mathbb{S}^5$ has been discussed in \cite{Bobev:2017asb}, and a powerful formula was conjectured for general $d = 2k+1$ from the point of view of high-derivative gravity \cite{Bueno:2018yzo}:
\be 
F''_d(\epsilon)\Big|_{\epsilon = 0} = \frac{(-1)^{\frac{d-1}{2}}\pi^{d+1}(d-1)^2 }{2d!}C_T.
\label{Prediction}
\ee 
Using the embedding map of $\mathbb{S}^7$ described in (\ref{S7Embedding}), we perform the integral as in \cite{Bobev:2017asb} for generic values of $d = 2k+1$ and find the result consistent with the prediction above, proving it from the field theory side, which is again a universal result applying for general CFTs living on squashed $(2k+1)$-spheres. This is the first time to obtain them from the field theory side up as far as we know. We put the detailed analysis of the integral in Appendix \ref{FppForGeneralk} for people interested.

Taking some special values $d = 3, 5, 7$, one gets:
\be 
F''_3(0) = - \frac{\pi^4}{3}C_T,\quad F''_5(0) = \frac{\pi^6}{15}C_T,\quad F''_7(0) = - \frac{\pi^8}{280} C_T.
\label{FppU1Bundle}
\ee 
The first two are identical to the results in \cite{Bobev:2017asb}. Comparing $F''(\epsilon)$ on the seven spheres with U(1) bundle in (\ref{FppU1Bundle}) and SU(2) bundle in (\ref{FppSU2Bundle}), we find the one with SU(2) bundle is twice larger, which means the squashing with SU(2) bundle perturbs the field theory more strongly, which is intuitively correct. 

\subsection{Conformally-coupled scalar}
The conformally coupled scalar theory, or the $O(N)$ model,\footnote{ In this paper, whenever we refer to the $O(N)$ model, we always assume $N = 1$, which is the conformally coupled scalar model. } is holographic dual to higher spin gravity \cite{Klebanov:2002ja}, whose free energy is different from the Einstein gravity that we studied. However, by comparing the free energies of Einstein's gravity and field theories here, we can observe some universal properties shared among them. \cite{Hartnoll:2005yc,Anninos:2012ft,Anninos:2013rza,Bobev:2016sap,Bobev:2017asb}. It would also be interesting to study the high spin gravity in the future. 

The partition function of massless conformally-coupled scalars living on squashed seven-sphere is as follows\footnote{The coefficient of the coupling term for general $d$ is $\frac{d-2}{4(d-1)}$. }
\be
Z_{\rm sc} = \int D\phi e^{-S_{\rm sc}[g_{ab},\phi] },\quad S_{\rm sc} = \frac{1}{2} \int d^7x\sqrt{g}  \left((\partial\phi)^2 + \frac{5}{24}R\phi^2\right).
\ee
After a Gaussian integration, the free energy is given by
\be
F_{\rm sc} = -\log Z_{\rm sc}  =  \frac{1}{2} \log \det \left(-\nabla^2 + \frac{5}{24}R\right).
\ee
The eigenvalues of Laplacian on the squashed seven sphere (\ref{SquashedSphereMetric1}) and their degeneracies are: \cite{Nilsson:1983ru}
\be
\begin{aligned}
	& \lambda_{n,r} = n(n+6) + \frac{1-\lambda^2}{\lambda^2} (n-2r)(n-2r+2),\quad\quad n = 0,1,2,\cdots; \\
	& m_{n,r} = \frac{1}{6}(n+3)(n-r+2)(n-2r+1)^2(r+1),\quad \ r = 0, 1, \cdots , \left[\frac{n}{2}\right]. \\
\end{aligned}
\label{ScalarSpectrum}
\ee
As shown in (\ref{RicciTensor}), the Ricci scalar of the two squashed sphere metrics are identical, thus the calculation here applies in both cases. We can obtain the spectrum of conformal Laplacian on squashed seven sphere with the Ricci scalar
\be
\tilde{\lambda}_{n,r} = \lambda_{n,r} + \frac{5}{24}R,\qquad R = 6 \left(8-2 \lambda ^2+\frac{1}{\lambda ^2}\right).
\ee
From (\ref{ScalarSpectrum}), we find the eigenvalues are unbounded, thus the free energy has a UV divergence and needs to be regularized. In the following, we will use two regularization methods. The first one is heat-kernel regularization following  \cite{Anninos:2012ft,Anninos:2013rza, Bobev:2016sap, Bobev:2017asb}, for which we can calculate $F_{\rm sc}(\lambda)$ numerically for general $\lambda$. The second one is zeta-function regularization following \cite{Hartnoll:2005yc, Bueno:2020odt}, with which we can calculate the derivatives at $\lambda = 1$ analytically. Combining the numerical results for generic $\lambda$ and the analytical results at $\lambda= 1$, we can justify the precision of numerical simulation, as well as compare them to the bulk calculation from various aspects.

\subsubsection{Heat-kernel regularization}

In this section, we follow the heat-kernel technique of \cite{Hawking:1976ja}, see also \cite{Monin:2016bwf, Vassilevich:2003xt}. Consider a general spectrum $\{\lambda_i\}$ with degeneracy $\{ m_i\} $, one can define the heat-kernel function and spectral zeta-function:
\be 
K(t) \equiv \sum_{i} m_i e^{-t\lambda_i},\quad \zeta_{\Delta}(p) \equiv \sum_{i}\frac{m_i}{\lambda_i^p}.
\ee 
After evaluating the Gaussian integral in the partition function, the free energy is proportional to the determinant of the kinetic operator. For conformally coupled scalar fields, we have
\be 
F = \frac{1}{2} \log\det \Delta = \frac{1}{2} \sum_i m_i \log \lambda_i = - \frac{1}{2} \zeta_\Delta'(0).
\label{FreeEnergySC}
\ee 
The spectral zeta function is related to the heat kernel with a Mellin transformation:
\be 
G(p) \equiv \Gamma(p)\zeta_\Delta(p) = \int_0^\infty dt K(t) t^{p-1}.
\ee 
On the left hand side above, we can expand $G(p)$ at small $p$:
\be 
G(p) = \frac{\zeta_\Delta(0)}{p} - \gamma\zeta_\Delta(0) + \zeta_\Delta'(0) +O(p).
\ee 
Meanwhile, on the right hand side, the integral is divergent in the small-$t$ region because of the divergence of heat kernel at small-$t$, which in general $d$-dimensional field theories is
\be 
K(t) =\sum_{k=0}^{\frac{d+1}{2}} a_{d/2 - k}t^{-d/2 + k} + O(t).
\ee 
In fact, both the divergence of $K(t)$ at $t = 0$ on the right hand side and the pole at $p = 0$ on the left hand side reflect the UV divergence of the determinant, which requires a regularization. The essence of heat-kernel regularization is to regulate the integral on the right hand side. Practically, we divide the integral domain into $(0,1] \cup [1,\infty)$ and perform the divergent integral over $(0,1]$ as follows, this essentially eliminates the divergence by throwing off the diverging contribution from a small cutoff:
\be \begin{aligned}
	&\int_0^1 dt \ \frac{1}{t} \ a_{d/2 - k}t^{-d/2 + k} = \lim_{\epsilon\rightarrow 0} \int_\epsilon^1 dt \ \frac{1}{t} \ a_{d/2 - k}t^{-d/2 + k}\\
	&\qquad\qquad\qquad  = \frac{ a_{d/2 - k}}{k-d/2} \lim_{\epsilon\rightarrow 0} \left(1 - \frac{1}{\epsilon^{d/2-k}} \right) \quad \Rightarrow \quad  \frac{ a_{d/2 - k}}{k-d/2}.
\end{aligned}\ee
After heat-kernel regularization, $G(p)$ is reduced to the following finite value:
\be 
G(p) = \int_0^1 dt\left[ K(t) - \sum_{k=0}^{\frac{d+1}{2}} a_{d/2 - k}t^{-d/2 + k} \right] t^{p-1} + \sum_{k=0}^{\frac{d+1}{2}} \frac{ a_{d/2 - k}}{k-d/2} + \int_1^\infty dt K(t) t^{p-1}.
\ee 
The expression above has no pole at $p = 0$, which requires $\zeta_\Delta(0) = 0$, thus the free energy (\ref{FreeEnergySC}) is equal to the integral of heat-kernel, this is the central equation we're going to calculate:
\be 
F_{\rm sc} = - \frac{1}{2}\zeta_\Delta'(0) = G(0) = -\frac{1}{2}  \int_0^\infty \frac{1}{t}K(t)dt.
\ee 
From the discussion above, the heat kernel regularization is essentially also a zeta function regularization through the spectral zeta function. Since the integrals above are evaluated numerically, we don't require $\lambda$ to be special values. For the conformally coupled scalars, we obtain the following coefficients of the divergent terms:
\be \small 
\begin{aligned}
	a_{7/2} =& \frac{\sqrt{\pi } \lambda ^3}{384}, \\
	a_{5/2} =& \frac{\sqrt{\pi } \lambda  \left(2 \lambda ^4-8 \lambda ^2-1\right)}{1536}, \\
	a_{3/2} =& \frac{\sqrt{\pi } \left(4 \left(255 \lambda ^6-504 \lambda ^4+257 \lambda ^2+60\right) \lambda ^2+15\right)}{184320 \lambda }, \\
	a_{1/2} =& -\frac{\sqrt{\pi } \left(2 \left(51100 \lambda ^{10}-60240 \lambda ^8-13674 \lambda ^6+6224 \lambda ^4+11109 \lambda ^2+1260\right)
		\lambda ^2+105\right)}{15482880 \lambda ^3}, \\
	a_{-1/2} =& \frac{\sqrt{\pi }}{247726080 \lambda ^5} [ 8 \left(733586 \lambda ^{14}-2142496 \lambda ^{12}+2212060 \lambda ^{10}-905552 \lambda ^8 \right. \\
	& \left. +81835 \lambda ^6+16808 \lambda ^4+5607
	\lambda ^2+420\right) \lambda ^2+105 ]. \\
\end{aligned}
\ee
\begin{figure}[t]
	\centering
	\includegraphics[width=8cm]{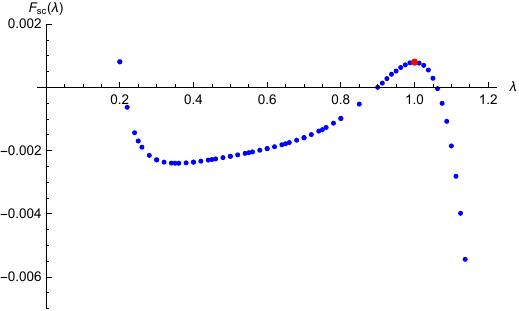}
	\caption{\rm Renormalized free energy with squashing parameter $\lambda$, the red points correspond to the free energy on the round sphere.}
	\label{Tbllambda}
\end{figure}
The calculation can theoretically be conducted using the formula listed above, but in practice, the double-summation of $n,r$ converges extremely slowly. So here we followed the technique of \cite{Anninos:2012ft}\footnote{The method has been nicely introduced in several places such as \cite{Anninos:2012ft, Bobev:2016sap, Bobev:2017asb}, thus we only present our final results here. }, using the Euler-MacLaurin formula to approximate the summation over thousands of points by integrals and function values, reducing the time-cost to an acceptable range. Using heat-kernel regularization, we numerically reproduced the round sphere free energy obtained in \cite{Klebanov:2011gs}; and the dependence of $\lambda$ is plotted in Fig.\ref{Tbllambda}. By comparing it with the bulk free energy shown in Fig.\ref{FscLambda}, we can identify similar behaviors as well as differences. The difference can mostly be attributed to the distinctions between strongly coupled holographic field theories and free ones we consider here, but also to field theory details such as correlation functions.

Using heat kernel regularization, we are able to evaluate not only the values of free energy but also derivatives with regard to the squashing $\lambda$. We studied the second and third-order derivatives at $\lambda = 1$, which provides a reference for our analytical results obtained in zeta function regularization. The procedure is nearly the same after taking the derivative of $\lambda$:
\be
\frac{d^n}{d\lambda^n} F_{\rm sc} = -\frac{1}{2}\int_0^\infty \frac{1}{t} \frac{\partial^n}{\partial \lambda^n}K(t) dt.
\ee
We also need to distill the divergent part of the integrand:
\be
\frac{\partial^n}{\partial\lambda^n}K(t) = \sum_{k=0}^{\frac{d+1}{2}} a_{d/2-k}^{(n)}t^{-d/2+k} + O(t).
\ee
By dividing the integral domain into $(0,1]$ and $[1,\infty)$, regulating the divergences, and evaluating the converging part numerically, we obtain the renormalized derivative, which is consistent with the results from zeta-function regularization which we obtain later:
\be
\begin{aligned}
	F_{\rm sc}^{(2)}(1) =& -0.289411 = 1.0009\left(-\frac{15\pi^2}{512}\right),\\
	F_{\rm sc}^{(3)}(1) =& -5.275 = 1.005\left(-\frac{62815 \pi ^2}{118272}\right).
\end{aligned}
\ee
For $F_{\rm sc}^{(2)}(1)$ and $F_{\rm sc}^{(3)}(1)$ at $\lambda = 1$, the divergent coefficients are as follows:
\be \small 
\begin{aligned}
	a_{7/2}^{(2)} =& \frac{\sqrt{\pi }}{64},\qquad a_{5/2}^{(2)} = -\frac{\sqrt{\pi }}{192},\qquad a_{3/2}^{(2)} = \frac{1453 \sqrt{\pi }}{30720}, \\
	a_{1/2}^{(2)} =& -\frac{152689 \sqrt{\pi }}{1290240},\qquad a_{-1/2}^{(2)} = \frac{2317181 \sqrt{\pi }}{41287680}. \\
	a_{7/2}^{(3)} =& \frac{\sqrt{\pi }}{64},\qquad a_{5/2}^{(3)} = \frac{3 \sqrt{\pi }}{64},\qquad a_{3/2}^{(3)} = \frac{16553 \sqrt{\pi }}{30720},\\
	a_{1/2}^{(3)} =& -\frac{2051699 \sqrt{\pi }}{1290240},\qquad a_{-1/2}^{(3)} = \frac{10868771 \sqrt{\pi }}{5898240}.\\
\end{aligned}
\ee

\subsubsection{Zeta-function regularization}
For the scalar spectrum (\ref{ScalarSpectrum}) we have
\be
F_{\rm sc}(\lambda) = \frac{1}{2} \log \det \left(-\nabla^2 + \frac{5}{24}R\right) = \frac{1}{2}\sum_{n=0}^\infty\sum_{r=0}^{\left[\frac{n}{2}\right]}m_{n,r}\log\tilde\lambda_{n,r}.
\ee
Thus, following \cite{Bueno:2020odt}, by formally commuting the infinite sum and derivative over $\lambda$, we obtain the following expressions for the $i$-th order derivative of $F_{\rm sc}$ at $\lambda = 1$:
\be
\begin{aligned}
	F^{(i)}_{\rm sc}(1) & =  \frac{1}{2}\sum_{n=0}^\infty\sum_{r=0}^{\left[\frac{n}{2}\right]}m_{n,r}\frac{d^i}{d\lambda^i}\log\tilde\lambda_{n,r}\Big|_{\lambda = 1} \\
	& = \frac{1}{2}\sum_{k=0}^\infty\sum_{r=0}^{k}m_{2k,r}\frac{d^i}{d\lambda^i}\log\tilde\lambda_{2k,r}\Big|_{\lambda = 1} + \frac{1}{2}\sum_{k=0}^\infty\sum_{r=0}^{k}m_{2k+1,r}\frac{d^i}{d\lambda^i}\log\tilde\lambda_{2k+1,r}\Big|_{\lambda = 1}. \\
\end{aligned}
\ee
For the $i = 0$ case, we met with a divergent series with logarithms, which needs to be regularized using the generalized zeta-function:
\be
\sum_{k=0}^\infty (k+ a)^n \log (k+a) = - \zeta_{a}'(-n),\quad \zeta_a(s) \equiv \sum_{m=0}^\infty \frac{1}{(a+m)^s},\quad a\in \mathbb{R}.
\label{GenerlizedZetaFunction}
\ee
The result we obtain, which is also shown in Table \ref{FpTbl}, is the same as in \cite{Klebanov:2011gs}:
\be
F_{\rm sc}(1) = \frac{60 \pi ^6 \log 2 + 82 \pi ^4 \zeta (3)-150 \pi ^2 \zeta (5)-945 \zeta (7)}{61440 \pi ^6} \approx 0.000797.
\ee
For the first-order derivative, which should vanish because of the vanishing of the conformal anomaly in odd-dimensional field theory, the expression is
\be
F_{\rm sc}'(1) = -\sum_{k=0}^\infty \frac{16}{21}\left(k + \frac{1}{2}\right) (k+1) \left(k +\frac{3}{2} \right)^2 ( k+2)\left(k+\frac{5}{2}\right).
\ee  
This expression seems to be non-zero using normal zeta-function regularization, contradicting the conformal symmetry. However, using a new parameter $k' = k + 3/2$, we get a non-trivial cancellation among generalized zeta functions:
\be
F_{\rm sc}'(1) =  -\frac{4}{21}\sum_{k'} (k'^2 - 5 k'^4 + 4 k'^6) =  -\frac{4}{21}(\zeta_{3/2}(-2) - 5\zeta_{3/2}(-4) + 4\zeta_{3/2}(-6)) = 0.
\ee
Since this kind of divergent series can't be regularized using a method that is both stable and linear, the shift of the summing parameter will bring a different answer. \cite{Monin:2016bwf} To guarantee we have the correct shift, one thing we can do is find a physical meaning of the parameter or use another summation method, such as comparing our results against those obtained by heat kernel regularization, integrated correlators, or the bulk results. Indeed, all these results justify that the universal value is obtained by omitting the term without the $\pi$ factors, which we will see below. Another justification is that although shifting variables $k' = k+a$ shifts the result, all terms with $\pi$ factors don't change.

For the second and third-order derivatives, we arrange the sum in the form where the first line is divergent and the second line is convergent:
\be \small 
\begin{aligned}
	F_{\rm sc}^{(2)} (1)& = \frac{1}{2} \sum_{k=0}^\infty  \left(\frac{544 k^6}{945}+\frac{272 k^5}{45}+\frac{23588 k^4}{945}+\frac{6928 k^3}{135}+\frac{17018 k^2}{315}+\frac{7003 k}{270}+\frac{6917}{1680} \right. \\
	&\qquad\qquad \left. -\frac{5 (64 k (2 k+7) (13 k (2 k+7)+155)+29619)}{48 (4 k+5)^2 (4 k+7)^2 (4 k+9)^2} \right) \\
	& = \frac{6499}{43200} + \frac{652-75 \pi ^2}{2560} \quad \Rightarrow \quad  -\frac{15 \pi ^2}{512}.
\end{aligned}
\ee
\begin{equation*} \small 
\begin{aligned}
	& F^{(3)}_{\rm sc}(1) \\
	&=  \frac{1}{2}\sum_{k=0}^\infty \left( -\frac{15296 k^6}{10395}-\frac{7648 k^5}{495}-\frac{234104 k^4}{3465}-\frac{233872
		k^3}{1485}-\frac{102317 k^2}{495}-\frac{143801 k}{990}-\frac{6209419}{166320} \right.\\
	&\left. -\frac{8 k (2 k+7) (8 k (2 k+7) (11960664 k (2
		k+7)+205534087)+9393298205)+142669238625}{18480 (4 k+5)^3 (4 k+7)^2 (4 k+9)^3}\right) \\
	&\qquad  =  -\frac{614077}{950400}-\frac{62815 \pi ^2}{118272} \quad \Rightarrow \quad -\frac{62815 \pi ^2}{118272}.\\
\end{aligned}\end{equation*} 
  The procedure outlined above can be directly applied to evaluate higher-order derivatives, which are less insightful from the holography point of view, thus we will not bother to put them here.
\begin{table}[!h]
	\renewcommand{\arraystretch}{1.7}
	\centering
	\begin{tabular}{cccc}
		\hline
		& $F$ & $F''$ & $F'''$ \\\hline
		Scalar  & $\frac{60 \pi ^6 \log 2 + 82 \pi ^4 \zeta (3)-150 \pi ^2 \zeta (5)-945 \zeta (7)}{61440 \pi ^6}$ & $-\frac{15 \pi ^2}{512}$ & $-\frac{62815 \pi ^2}{118272}$ \\
		Fermion & $\frac{300 \pi ^6 \log2 + 518 \pi ^4 \zeta (3)+1050 \pi ^2 \zeta (5)+945 \zeta (7)}{7680 \pi ^6}$ & $-\frac{45 \pi ^2}{64}$ & $-\frac{101335 \pi ^2}{7392}$ \\\hline
	\end{tabular}
	\caption{\rm $F^{(i)}_{\rm sc}(1)$ and $F^{(i)}_{\rm f}(1)$ for $i=0,2,3$.}
	\label{FpTbl}
\end{table}

\subsection{Free fermion}
In this section, we calculate the derivatives of free energy at $\lambda = 1$ for free fermions analytically, using zeta function regularization introduced before. The free energy is
\be
F_{\rm f}(\lambda) = - \log \det \left(-i\slashed{\nabla}\right) = -\sum_{n,r}m_{n,r}\log \tilde{\lambda}_{n,r}.
\ee
The spectrum of Dirac operator on squashed seven-sphere with SO(5)$\times$SO(3) isometry was calculated in \cite{Nilsson:1983ru}, and in \cite{Eastaugh:1985ew} by another method. The eigenstates of Dirac operator in $\mathbb{S}^7$ correspond to the SO(8) irreps $(n,0,0,1)$ and $(n,0,1,0)$, with negative and positive values respectively. \cite{Duff:1986hr} Consider the branching rules of the two SO(8) irreps under SO(5)$\times$SO(3):
\be \small 
\begin{aligned}
	(n,0,0,1) & \rightarrow \sum_{r=0}^{[n/2]} (n+1-2r,r;n+1-2r) + \sum_{r=0}^{[n/2]}(n+1-2r,r;n-1-2r)\\
	& + \sum_{r=0}^{[n/2]}(n-1-2r, r+1;n+1-2r) + \sum_{r=0}^{[n/2]} (n-1-2r,r;n-1-2r). &\\
	(n,0,1,0) & \rightarrow \sum_{r=0}^{[n/2]+1} (n-2r,r+1;n-2r) + \sum_{r=0}^{[(n-1)/2]} (n-2r,r;n-2r) \\
	& + \sum_{r=0}^{[n/2]} (n-2r,r;n-2r+2) + \sum_{r=0}^{[n/2-1]} (n-2r,r; n-2r-2). \\
\end{aligned}
\label{decomposition}
\ee 
The label $(p, q; r)$ correspond to irrep of SO(5) with Dynkin indices $(p, q)$ and irrep for Sp(1) with index $r$, whose dimensions are:
\be 
\begin{aligned}
	&m_{p,q} = \frac{1}{6}(p+1)(q+1)(p+q+2)(p+2q+3),\quad m_r = r + 1.  \\
\end{aligned}
\ee 
Thus the degeneracy of irrep $(p, q; r)$ is $m_{p,q,r} = m_{p,q}m_r$. In the branching rule, there're three classes of SO(5)$\times$SO(3) irreps, with specific relations between $r$ and $p$: (i)$(p,q;p)$, (ii)$(p,q;p+2)$, (iii)$(p,q;p-2)$. 

For case (i), the eigenvalues $\tilde\lambda$ are solutions of the following quatic equation:\footnote{Notice that the equation in \cite{Nilsson:1983ru} has a typo in the last term in the denominator.}
\be 
\begin{aligned}
	& (\tilde\lambda-\sigma_{1+})(\tilde\lambda-\sigma_{1-})(\tilde\lambda-\sigma_{2+})(\tilde\lambda-\sigma_{2-}) - \frac{\left(\lambda ^2-1\right) \left(5 \lambda ^2-1\right)}{\lambda^4}  p (p+2) = 0; \\
	&\qquad  \sigma_{1\pm} = -\frac{3}{2}\lambda \pm  \frac{1}{2\lambda}\sqrt{9\lambda^4 + 2\lambda^2(8 p q+8 p+8 q^2+24 q+9) + 4 p^2+8 p+9},\\
	&\qquad \sigma_{2\pm} = \frac{\lambda}{2} \pm \frac{1}{2\lambda}\sqrt{\lambda^4 +2\lambda^2(8 p q+8 p+8 q^2+24 q+17) + 4 p^2+8 p+1}.
\end{aligned}
\ee 
The equation becomes trivial when $\lambda^2 = 1, 1/5$. And we will name the four corresponding branches of eigenvalues $(\Sigma^{1+}, \Sigma^{2+},\Sigma^{1-},\Sigma^{2-})$, which reduce to $(\sigma^{1+}, \sigma^{2+},\sigma^{1-},\sigma^{2-})$ respectively if we take $\lambda^2 = 1, 1/5$.

For cases (ii) and (iii), the eigenvalues are denoted by $D^{2\pm}$ and $D^{3\pm}$ respectively:
\be
\begin{aligned}
	& D^{2\pm} = \frac{\lambda}{2} \pm \frac{1}{2\lambda}\sqrt{\lambda^4+2\lambda^2(8 p q+2 p+8 q^2+24 q+5)+(2p+5)^2}; \\
	& D^{3\pm} = \frac{\lambda}{2} \pm \frac{1}{2\lambda}\sqrt{\lambda^4+2\lambda^2(8 p q+14 p+8 q^2+24 q+17)+(2p-1)^2}. \\
\end{aligned}
\ee
To determine which eigenvalue to choose for the irreps in (\ref{decomposition}), we need to compare the above eigenvalues with \cite{Duff:1986hr}, which discussed the special case $\lambda^2 = 1/5$. It turns out that corresponding to the eight irreps of SO(5)$\times $SO(3) appearing in the branching rule following the order of \eqref{decomposition}, we need to specify the following eigenvalues:
\be
\Sigma^{2-},\ D^{3-},\ D^{2-},\ \Sigma^{1-};\qquad \Sigma^{1+},\ \Sigma^{2+},\ D^{2+},\ D^{3+}.
\ee
When $\lambda = 1$, the eigenvalues and degeneracies above are significantly simplified:
\be
\begin{aligned}
	D^{+} &= n + \frac{7}{2},\quad D^{-} = -n - \frac{7}{2},\quad n=0,1,2,...;\\
	m_n &= \frac{1}{90} (n+1) (n+2) (n+3) (n+4) (n+5) (n+6). \\
\end{aligned}
\ee
To calculate $F_{\rm f}(1)$, we need generalized zeta function as in (\ref{GenerlizedZetaFunction}) to regulate logarithmic terms, which is also shown in Table \ref{FpTbl}, and corresponds to the results of \cite{Klebanov:2011gs}:
\be
F_{\rm f}(1) = \frac{300 \pi ^6 \log2 + 518 \pi ^4 \zeta (3)+1050 \pi ^2 \zeta (5)+945 \zeta (7)}{7680 \pi ^6} \approx 0.0369.
\ee
The first, second, and third-order derivatives can be calculated using the ordinary zeta-function regularization, and we extract the terms with factors of $\pi$ as the physically relevant ones as argued before:
\begin{equation*}  \small 
\begin{aligned}
	F'_{\rm f}(1) = & \frac{128}{105}\sum_{k=0}^\infty (k+1)\left(k+\frac{3}{2}\right)(k+2)\left(k+\frac{5}{2}\right)(k+3)\\
	= &\frac{32}{105} \sum_{m=2}^\infty (4m^6 - 5m^4 + m^2) = \frac{32}{105}[4\zeta(-6)-5\zeta(-4)+\zeta(-2)] = 0.\\
	F''_{\rm f}(1) = & -\sum_{k=0}^\infty\left(\frac{2176 k^6}{945}+\frac{8704 k^5}{315}+\frac{125984 k^4}{945}+\frac{311552
		k^3}{945}+\frac{3100 k^2}{7}+\frac{295696 k}{945}+\frac{112481}{1260}\right.\\
	& \left.  -\frac{256}{105 (2 k+5)}+\frac{45}{8 (4 k+7)^2}+\frac{45}{8 (4 k+9)^2}+\frac{256}{105 (2
		k+3)}\right)\\
	= & -\frac{5608873}{264600} + \left( \frac{14269}{2520} - \frac{45 \pi ^2}{64}\right) \quad \Rightarrow \quad - \frac{45 \pi ^2}{64}.\\
	F'''_{\rm f}(1) =& - \sum_{k=0}^\infty\left(\frac{16 k (k+4) (8 k (k+4) (956 k (k+4)+9981)+255129)+5118283}{20790}\right.\\
	&\qquad\qquad \left. + \frac{\frac{600000}{2 k+5}-\frac{506675}{(4 k+7)^2}-\frac{506675}{(4
			k+9)^2}-\frac{248832}{(2 k+5)^3}-\frac{600000}{2 k+3}+\frac{248832}{(2 k+3)^3}}{4620}\right)\\
	=& - \frac{101335 \pi ^2}{7392} + \frac{54409409}{363825}\quad \Rightarrow \quad - \frac{101335 \pi ^2}{7392}.
\end{aligned}
\end{equation*}

\section{Comparison}
\label{SectionofComparison}
Up to now, we have studied the free energies of three theories: holographic CFT corresponding to Euclidean Einstein gravity, O($N$) vector model, and free fermion model. In this section, we will compare their free energies as functions of the squashing parameter $\lambda$. 

 The number of degrees of freedom in field theory is characterized by its central charge $C_T$ in the stress-tensor two-point functions (\ref{StressTensor2ptFunction}), thus to compare the free energy among different theories, the most convenient normalization is to divide them by the corresponding central charges. For free scalar and massless Dirac fermion, the central charges are given by, respectively: \cite{Bueno:2020odt,Osborn:1993cr,Erdmenger:1996yc,Buchel:2009sk}\footnote{In \cite{Bobev:2017asb}, the convention is different, where the factor of $V_{\mathbb{S}^{d-1}}$ is absent. }
\be
C_{T, {\rm sc}} = \frac{d}{d-1}\frac{1}{V_{\mathbb{S}^{d-1}}^2},\qquad C_{T,{\rm f}} = \frac{d}{2}2^{[d/2]}\frac{1}{V_{\mathbb{S}^{d-1}}^2},\quad V_{\mathbb{S}^{d-1}} \equiv \frac{2\pi^{d/2}}{\Gamma\left( d/2\right)} .
\ee
The two-point function of stress tensor can also be evaluated holographically from the bulk side, and the value of $C_T$ in Einstein gravity is given by \cite{Buchel:2009sk, Liu:1998bu}:
$$
C_{T, {\rm holo}} = \frac{\Gamma(d+2)}{8(d-1)\Gamma(d/2)\pi^{(d+2)/2}} \frac{\ell^{d-1} }{G_N} .
$$
Specifically for $d=7$, we have:
$$
C_{T, {\rm sc}} = \frac{525}{512\pi^6},\quad C_{T, {\rm f}} = \frac{1575}{64\pi^6}, \quad C_{T,{\rm holo}} = \frac{448}{\pi^5} \frac{\ell^6}{G_N}.
$$
\begin{figure}[t]
	\centering
	\includegraphics[width=8cm]{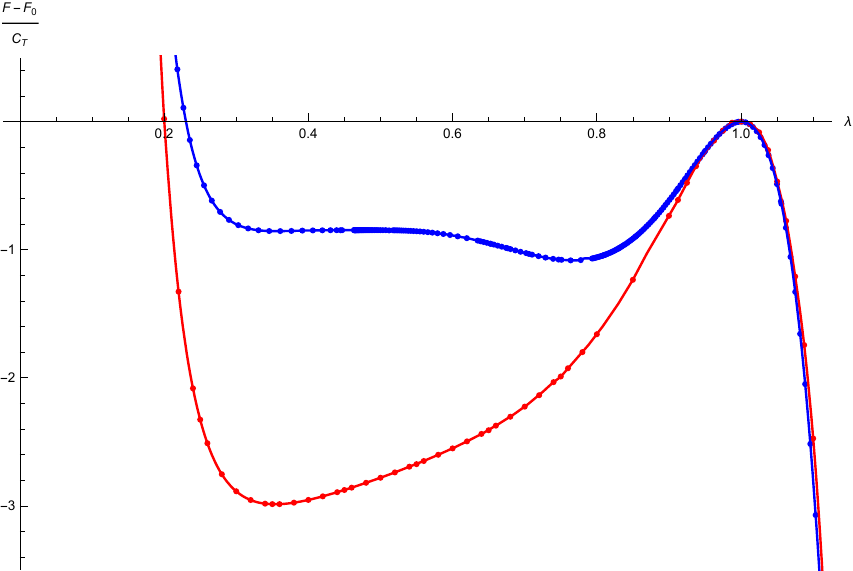}
	\caption{\rm Blue points: gravitational free energy. Red points: conformally coupled scalar action. The zero-point energy where $\lambda = 1$ is extracted, and $C_T$ is divided. The lines are interpolation lines to make the plots more clear.}
	\label{Comparison}
\end{figure}
The first comparison can be made between normalized $F_{\rm bulk}$ and $F_{\rm sc}$ for general $\lambda$ collected in Fig. \ref{Comparison}. Except for the expected correspondence at weak and strong deformation regions, they seem to behave differently, especially around $\lambda_* = 1/\sqrt{5} \approx 0.447$. There appear to be some local minima in both cases which are less understood. The gravitational free energy seems to have a platform around $\lambda_*$, which deserves further numerical study with higher precision.

In the weak-squashing regime, we collect the analytic results of second-order and third-order derivatives at $\lambda=1$ for scalar and fermion in Table \ref{CompareTbl}. In the same table, we also include the same quantities of $F_{\rm bulk}$ obtained by numerical fitting. The small discrepancy of the second-order derivative is probably due to numerical errors, which reflects our normalization by dividing $C_T$ in each theory is reasonable. The three-point function of the stress tensor depends on the details of the field theory, which is different in different theories. It was observed in \cite{Bobev:2017asb} that for some special field theories on U(1) squashed three spheres, the third-order derivatives of free energies of the squashing parameter $\epsilon$ appear to be vanishing, which is no longer true in our example. Take the holographic field theory as an example, we have:
\be 
    \frac{1}{C_T} \frac{d^3 F}{d\epsilon^3}\Big|_{\epsilon = 0} = \frac{1}{8C_T} F'''(1) - \frac{3}{8C_T} F''(1) \approx -670,
\ee 
which is different from zero. It will be interesting to explore this further.

\begin{table}[!h]
	\centering
	\renewcommand{\arraystretch}{1.5}
	\begin{tabular}{cccc}
		\hline & scalar & free fermion & bulk\\\hline
		$ F''(1)/C_T$ &  $-\frac{\pi^8}{35}\approx -271.1$ & $-\frac{\pi^8}{35}\approx -271.1$ & $-271.8$ \\
		$ F'''(1)/C_T$ & $-\frac{12563 \pi ^8}{24255}\approx -4900$ & $-\frac{40534 \pi ^8}{72765}\approx -5300$ & $-6200$\\\hline
	\end{tabular}
	\caption{\rm Comparison of the second and third-order derivatives of free energy at $\lambda = 1$.}
	\label{CompareTbl}
\end{table}


\begin{figure}[t]
	\centering
	\includegraphics[width=7cm]{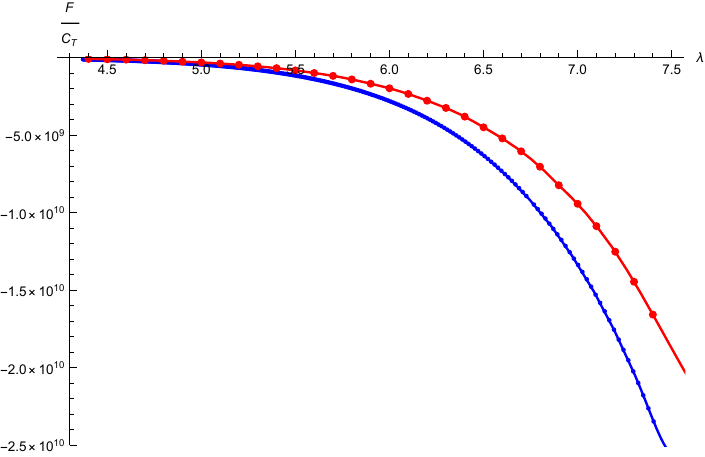}
	\qquad \qquad 
	\includegraphics[width=7cm]{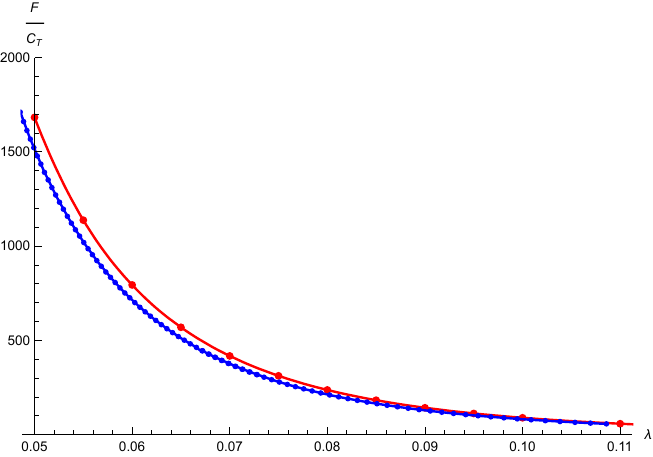}
	\hfil 
	\includegraphics[width=7cm]{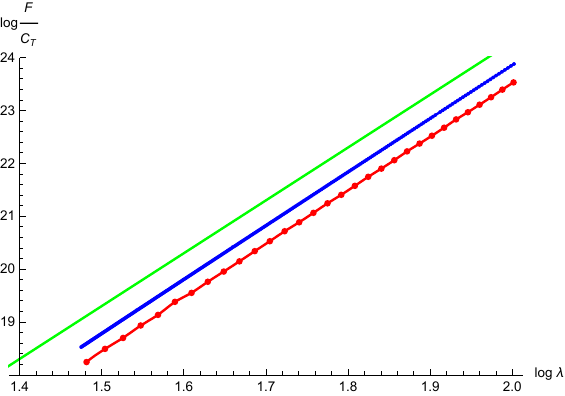}
	\qquad \qquad 
	\includegraphics[width=7cm]{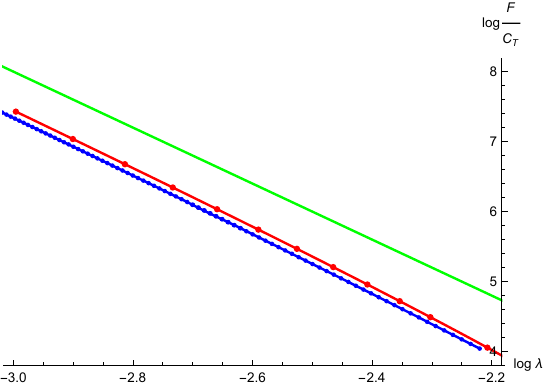}
	\caption{{\rm Free energy at large and small $\lambda$, respectively. The blue line is bulk free energy, and the red one is for free scalar. The green lines are reference lines, whose slopes are 10 and $-4$ respectively.}}
	\label{LargeSquashing}
\end{figure}
The comparison at strong-squashing regime, corresponding to big and small values of $\lambda$, has been reported in the literature. \cite{Aharony:2019vgs,Hartnoll:2005yc} Our results are collected in Fig. \ref{LargeSquashing}. Although the free energies are not exactly the same as in weak-squashing regime, they turn out to be proportional to each other and have the same scaling of $\lambda$ or $\lambda^{-1}$ respectively, which is the same as AlAdS$_4$ study in \cite{Hartnoll:2005yc}.

For small $\lambda$, the size of the SU(2) bundle can be omitted compared with the $\mathbb{S}^4$ base space, this means $b(r)$ will cap off before $a(r)$ as $r$ go from infinity, where the local geometry is given by $\mathbb{S}^4\times \mathbb{R}^4$, corresponding to a Bolt space. Following \cite{Aharony:2019vgs}, we expect the free energy of the conformal field theory to be identical to a theory living on $\mathbb{S}^4$, which is proportional to the normalized volume of $\mathbb{S}^4$ and scales like $F\sim \frac{1}{\lambda^4}$, and indeed they have been reproduced for free energies both in the bulk and boundary. 

At large $\lambda$, the $\mathbb{S}^4$ becomes tiny and the fiber effect becomes strong, and numerics suggests to us the power is $F\sim \lambda^{10}$. This power can be compared against \cite{Aharony:2019vgs}: there they consider an asymptotic boundary $\mathbb{S}^4\times\mathbb{S}^3$ there when $\mathbb{S}^3$ becomes large, the near horizon geometry becomes $\mathbb{R}^5\times \mathbb{S}^3$, corresponding to a CFT living on $\mathbb{S}^3$, thus the free energy will scale as $\lambda^3$, which is much different from our result. The difference in our case is that we don't have such a solution where $a(r)$ caps off before $b(r)$, thus the field theory argument above doesn't work anymore. Notice that $10 = 7 + 3$, which is the sum of the dimension of the full boundary and the bundle, it will be interesting to understand this scaling better.

We can also discuss the large squashing behavior for field theories living on the U(1) squashed $(2k+1)$-spheres. If we still label the size of the bundle as $\lambda$, then as $\lambda \rightarrow 0$, the free energy scales as an effective theory on $\mathbb{S}^{2k}$, thus $F\sim \lambda^{-2k}$; and as $\lambda \rightarrow \infty$, we expect a power equal to the sum of boundary dimension and the fibre, which is $\lambda^{2k+2}$. The special case $k = 1$ has been already considered in the literature. At the bulk side, it was done in \cite{Emparan:1999pm} where $O(\lambda^4)$ scaling at large $\lambda$ and $O(\lambda^{-2})$ at small $\lambda$ are obtained analytically. The $O(\lambda^{-2})$ scaling has also been reproduced for conformally coupled scalar and fermions. \cite{Hartnoll:2005yc, Bobev:2017asb, DeFrancia:2000xm}

\section{Discussion}

As the first result of this paper, we construct a novel one-parameter family of AlAdS$_8$ solutions of Euclidean Einstein gravity preserving SO(5)$\times$SO(3) isometry with two closely related metrics. We study the renormalized gravitational free energy as a function of the squashing parameter and show its correspondence with free field theories living on a squashed seven sphere. With the solutions containing SU(2) fiber bundles, our work has extended the literature on AdS vacuum solutions. On the field theory side, we have studied the free energy for an arbitrary CFT living on a perturbed metric. Using zeta-function and heat-kernel regularization, we study the free energy of free conformal scalar and free fermion theories living on the squashed sphere. We compare free energies in the bulk and boundary both at small and large deformations.

As a second result of this paper, we expressed the second-order derivative of the free energy with regard to the squashing parameter in terms of the central charge in general CFT. The relation is identical to the corresponding one in the bulk, which reflects the universal property of conformal symmetry and acts as a sanity check of our result. We checked this for general conformal field theories living on the U(1) squashed sphere or the SU(2) squashed seven spheres. From a renormalization group point of view, the squashing on the round sphere induces an RG flow generated by the stress tensor of the field theory, along which the $\widetilde F$ quantity related to the free energy must be decreasing, this is also confirmed in our analysis.

\subsection*{Future directions}

The story we uncovered is interesting but there are some open questions left by our analysis. Some of them are:

First, more investigations on the field theory free energy will further support our results. It's interesting to evaluate the renormalized free energy for free fermions using heat-kernel regularization to get the strong deformation behavior, where a crucial step is to obtain the heat-kernel coefficients. One should also try to obtain the strong deformation behavior analytically, following \cite{DeFrancia:2000xm}.

Secondly, there're several aspects of bulk calculation that are interesting to explore. We didn't study the detailed behavior of the renormalized free energy when the squashing parameter approaches the special value $\lambda_* = \frac{1}{\sqrt{5}}$, which requires higher numerical precision. It would be interesting to see whether there is an oscillating behavior around the conically singular solution as observed in \cite{Aharony:2019vgs}. This special limit can also be studied from the NUT side analytically following the metric perturbation method as in \cite{Aharony:2019vgs}.

Furthermore, our second metric ansatz turns out to be more general than the first one and admits three squashing parameters. The discussions in the appendix \ref{solutionAnsatz1} only show the behaviors of the metric, without exploring further the free energy. It would be definitely interesting to investigate the extended phase space of the AlAdS$_8$ solution as well as its holographic partners. We conjecture that there should be an interface between the NUT and Bolt phase in the bulk configuration space parametrized by $(\lambda_1, \lambda_2, \lambda_3 ) \equiv \left(\frac{\cF_{30}}{\cF_{20}}, \frac{\cF_{40}}{\cF_{20}}, \frac{\cF_{50}}{\cF_{20}} \right) $, passing through the special point corresponding to $\lambda_* = \frac{1}{\sqrt{5}}$. On the field theory side, we need to generalize the spectrum of conformal Laplacian \cite{Nilsson:1983ru, Eastaugh:1985ew} when the symmetry group is further broken from SO(5)$\times$SO(3) to SO(5)$\times$U(1). At strong deformation regime, we expect the bulk and boundary free energy to scale consistently as a homogeneous polynomial of $(\lambda_1, \lambda_2, \lambda_3)$ whose order is dictated by our analysis which essentially corresponds to the special case $\lambda_1 = \lambda_2 = \lambda_3$.

We also need to understand better the $\lambda^{10}$ scaling of the free energies, which reflects important properties of the field theory. Notice that a similar scaling also emerges for U(1) squashed spheres, a numerical study would also be helpful for those cases. We are also looking forward to a formal justification of our steps to evaluate integrated two-point correlators on the sphere, which may also help us evaluate three-point correlators of stress tensors.

Based on our work, there are also many generalizations that are worthy of exploring.

Given the lack of superconformal field theories in $d = 7$ \cite{Nahm:1977tg, Shnider:1988wh}, there have been many explorations on non-conformal supersymmetric Yang-Mills theory living on seven manifolds, which can be constructed as the worldvolume theory of the D6 branes in the Euclidean type IIA$^*$ theory. \cite{Prins:2018hjc} With supersymmetries, localization calculations can be done to extract information about the free energy and the Wilson loops expectation values. \cite{Bobev:2019bvq} With background fluxes turned on, we expect the free energy to be different from our case, but it would be interesting to compare our results and illustrate the difference.

As we mentioned in the beginning, as dictated by the Klebanov-Polyakov correspondence \cite{Klebanov:2002ja}, the O($N$) vector field and free fermion on the boundary are dual to higher spin gravity in the bulk. The higher spin modes in the latter contributes non-trivially to the free energy, and thus needs to be studied independently. There are two types of higher spin gravities, called A and B, correspond to complex scalar and free fermion fields on the boundary respectively.\footnote{See \cite{Giombi:2012ms} for a review and \cite{Giombi:2014iua,Giombi:2016pvg,Brust:2016xif} for results with AdS$_8$.} Following \cite{Giombi:2014iua, Giombi:2016pvg, Brust:2016xif} for a new solution similar to our ansatzes \eqref{Ansatz1} and \eqref{Ansatz2} will be desirable. This will provide new opportunities for quantitative checks of the higher-spin/vector model duality.

The authors of \cite{Ooguri:2016pdq} conjectured that any non-supersymmetric AdS vacua supported by fluxes must be unstable. An example with a 11D Euclidean AdS$_5 \times \mathbb{CP}^3$ backups the conjecture. \cite{Ooguri:2017njy} It would be interesting to study the stability under real-time evolution of our bulk space, following \cite{Bizon:2007zf} whose geometry is asymptotically flat.

One can further explore the integrated three-point functions of stress tensor on seven spheres, and compare them against the proposal in \cite{Bueno:2018yzo, Bueno:2020odt}. It was pointed out in \cite{Bobev:2017asb, Bueno:2018yzo} that free bosons and free fermions may play as lower and upper bounds for the free energies in general field theories. For U(1) bundled $(2k+1)$-spheres, this can be explicitly checked by numerically evaluating the renormalized Euclidean Einstein-Hilbert action as well as analytically on the field theory side using the zeta-function regularization with the known eigenvalues of Laplacian \cite{Bobev:2017asb} and Dirac operators \cite{Bar:1996wqp}.

Our metric ansatz is based on quaternionic line bundle, it might be interesting to also consider octoninic line bundle considered in \cite{Bakas:1998rt, Kanno:1999hr}. Generalizing our solutions here to include higher derivative corrections is also an interesting direction to explore. \cite{Bueno:2018uoy} A more ambitious plan is to follow the top-down approach of \cite{Bobev:2019bvq, Cordova:2018eba} to understand the AdS$_8$ vacua in string theory.

\section*{Acknowledgement}
We would like to thank Valentin Reys for his useful comments on zeta-function regularization, Pablo Cano, Pieter-Jan de Smet, and Kwinten Fransen for interesting discussions, and especially Nikolay Bobev for so much help he gave and carefully reading and giving comments on the draft. The useful comments from the referees of JHEP is also appreciated. X.Z. would also like to thank ITF of KU Leuven for the hospitality during his early visiting in 2019. The work of X.Z. is supported by the KU Leuven C1 grant ZKD1118 C16/16/005.

\appendix

\section{Partial results with the second ansatz}
\label{solutionAnsatz1}

In this section, we collect our results for the second squashed sphere metric (\ref{Ansatz2}). We solved the Einstein equations both perturbatively and numerically, and evaluate the initial values that can be integrated up to the boundary. The results indicate a richer family of solutions worth exploring further.

\subsection{Einstein equations of motion}
\label{EinsteinEOM}

Using the tetrad method introduced in the main text (\ref{TetradFormalism}), the Einstein equations for the metric (\ref{SquashedSphereMetric1}) are as follows:
\be 
\begin{aligned}
	&\frac{ f_1^2 f_3^2}{2 f_2^4}+\frac{ f_1^2 f_4^2}{2 f_2^4}+\frac{ f_1^2 f_5^2}{2 f_2^4}-\frac{6
		f_1^2}{f_2^2}+\frac{f_1^2 f_5^2}{4  f_3^2 f_4^2}+\frac{f_1^2 f_4^2}{4  f_3^2 f_5^2}		+\frac{f_1^2 f_3^2}{4 
		f_4^2 f_5^2}-\frac{f_1^2}{2  f_3^2}-\frac{f_1^2}{2  f_4^2} &\\ 
	&	-\frac{f_1^2}{2  f_5^2}-21 f_1^2+\frac{4 f_2' f_3'}{f_2
		f_3}+\frac{4 f_2' f_4'}{f_2 f_4}+\frac{4 f_2' f_5'}{f_2 f_5}+\frac{6 f_2'^2}{f_2^2}+\frac{f_3' f_4'}{f_3
		f_4}+\frac{f_3' f_5'}{f_3 f_5}+\frac{f_4' f_5'}{f_4 f_5} = 0, &\\
	&-\frac{3 f_1' f_2'}{f_1 f_2}-\frac{f_1' f_3'}{f_1 f_3}-\frac{f_1' f_4'}{f_1 f_4}-\frac{f_1' f_5'}{f_1
		f_5}-\frac{3 f_1^2}{f_2^2}+\frac{f_1^2 f_5^2}{4  f_3^2 f_4^2}+\frac{f_1^2 f_4^2}{4  f_3^2 f_5^2}
	+\frac{f_1^2 f_3^2}{4
		f_4^2 f_5^2}-\frac{f_1^2}{2  f_3^2}-\frac{f_1^2}{2  f_4^2}-\frac{f_1^2}{2  f_5^2}-21 f_1^2	&\\
	&	+\frac{3
		f_2''}{f_2}+\frac{3 f_2' f_3'}{f_2 f_3}+\frac{3 f_2' f_4'}{f_2 f_4}+\frac{3 f_2' f_5'}{f_2 f_5}+\frac{3
		f_2'^2}{f_2^2}
	+\frac{f_3''}{f_3}+\frac{f_3' f_4'}{f_3 f_4}+\frac{f_3' f_5'}{f_3 f_5} +\frac{f_4''}{f_4}+\frac{f_4'
		f_5'}{f_4 f_5}+\frac{f_5''}{f_5} = 0, &\\
	&	-\frac{4 f_1' f_2'}{f_1 f_2}-\frac{f_1' f_4'}{f_1 f_4}-\frac{f_1' f_5'}{f_1 f_5}+\frac{3  f_1^2 f_3^2}{2
		f_2^4}+\frac{ f_1^2 f_4^2}{2 f_2^4}+\frac{ f_1^2 f_5^2}{2 f_2^4}-\frac{6 f_1^2}{f_2^2}-\frac{f_1^2 f_5^2}{4  f_3^2
		f_4^2}
	-\frac{f_1^2 f_4^2}{4  f_3^2 f_5^2}  +\frac{3 f_1^2 f_3^2}{4  f_4^2 f_5^2} &\\
	&	+\frac{f_1^2}{2  f_3^2}-\frac{f_1^2}{2
		f_4^2}-\frac{f_1^2}{2  f_5^2}-21 f_1^2		+\frac{4 f_2''}{f_2}+\frac{4 f_2' f_4'}{f_2 f_4}+\frac{4 f_2' f_5'}{f_2
		f_5}+\frac{6 f_2'^2}{f_2^2}+\frac{f_4''}{f_4}+\frac{f_4' f_5'}{f_4 f_5} +\frac{f_5''}{f_5} = 0, &\\
	&	-\frac{4 f_1' f_2'}{f_1 f_2}-\frac{f_1' f_3'}{f_1 f_3}-\frac{f_1' f_5'}{f_1 f_5}+\frac{ f_1^2 f_3^2}{2
		f_2^4}+\frac{3  f_1^2 f_4^2}{2 f_2^4}+\frac{ f_1^2 f_5^2}{2 f_2^4}-\frac{6 f_1^2}{f_2^2}-\frac{f_1^2 f_5^2}{4  f_3^2
		f_4^2}
	+\frac{3 f_1^2 f_4^2}{4  f_3^2 f_5^2} 	-\frac{f_1^2 f_3^2}{4  f_4^2 f_5^2} &\\
	&-\frac{f_1^2}{2  f_3^2}+\frac{f_1^2}{2
		f_4^2}-\frac{f_1^2}{2  f_5^2}-21 f_1^2+\frac{4 f_2''}{f_2}+\frac{4 f_2' f_3'}{f_2 f_3}+\frac{4 f_2' f_5'}{f_2
		f_5}+\frac{6 f_2'^2}{f_2^2}+\frac{f_3''}{f_3}+\frac{f_3' f_5'}{f_3 f_5} +\frac{f_5''}{f_5}	= 0, &\\
	&	-\frac{4 f_1' f_2'}{f_1 f_2}-\frac{f_1' f_3'}{f_1 f_3}-\frac{f_1' f_4'}{f_1 f_4}+\frac{ f_1^2 f_3^2}{2
		f_2^4}+\frac{ f_1^2 f_4^2}{2 f_2^4}+\frac{3  f_1^2 f_5^2}{2 f_2^4}-\frac{6 f_1^2}{f_2^2}+\frac{3 f_1^2 f_5^2}{4 
		f_3^2 f_4^2}	-\frac{f_1^2 f_4^2}{4  f_3^2 f_5^2}	-\frac{f_1^2 f_3^2}{4  f_4^2 f_5^2}&\\
	&-\frac{f_1^2}{2 
		f_3^2}-\frac{f_1^2}{2  f_4^2}+\frac{f_1^2}{2  f_5^2}-21 f_1^2+\frac{4 f_2''}{f_2}
	+\frac{4 f_2' f_3'}{f_2 f_3}+\frac{4
		f_2' f_4'}{f_2 f_4}+\frac{6 f_2'^2}{f_2^2}+\frac{f_3''}{f_3}+\frac{f_3' f_4'}{f_3 f_4} +\frac{f_4''}{f_4} = 0, &\\
\end{aligned}
\label{EOMAnsatz1}
\ee
where we have taken the cosmological constant explicitly to be $-21$. Only four of the equations are independent because of Bianchi Identity. If one considers the special case where: 
\be
f_1(r) \rightarrow 1,\quad f_2(r)\rightarrow a(r),\quad f_3(r) = f_4(r) = f_5(r) \rightarrow b(r),
\ee
then the last three equations become identical, and the remaining three equations become identical to the three equations for our first metric ansatz in (\ref{EinsteinEquations3}). Thus the solutions discussed in the main text belong to a subset of solutions here, which we will solve both analytically at the near horizon and near boundary regions, and numerically for general squashing parameters $\lambda_1, \lambda_2, \lambda_3$.

\subsection{Large radius expansion}
We perform the following Fefferman-Graham expansion:
\be
f_1(r) = 1,\quad f_i(r) = \sum\limits_{j = 0}^{+\infty} \cF_{ij}e^{-(j-1)r},\quad i = 2,3,4,5
\ee
We solve the expansion perturbatively up to $O(\exp(-9r))$, it turns out the solutions are determined by totally 7 free parameters which can be chosen as $\{\cF_{i0}, \cF_{j7}\}_{i = 2,3,4,5}^{j = 2,3,4}$. The first several terms up to $O(e^{-3r})$ are
\be \small 
\begin{aligned}
	f_2(r) = &\cF_{20}e^{r} + e^{-r} \left(\frac{ \cF_{30}^2+\cF_{40}^2+\cF_{50}^2}{24
		\cF_{20}^3}\right.&\\
	&\left.-\frac{\cF_{20} \left(\cF_{30}^4-2
		\left(\cF_{40}^2+\cF_{50}^2\right)
		\cF_{30}^2+\left(\cF_{40}^2-\cF_{50}^2\right)^2\right)}{240 
		\cF_{30}^2 \cF_{40}^2 \cF_{50}^2}-\frac{1}{5 \cF_{20}}\right) + O(e^{-3r}), &\\
	f_3(r) = &\cF_{30}e^{r} + \frac{e^{-r}}{240\cF_{30}}\left( -\frac{2  \left(13 \cF_{30}^2+\cF_{40}^2+\cF_{50}^2\right)
		\cF_{30}^2}{\cF_{20}^4}\right.&\\
	&\left.	+\frac{-13 \cF_{30}^4+2
		\left(\cF_{40}^2+\cF_{50}^2\right) \cF_{30}^2+11
		\left(\cF_{40}^2-\cF_{50}^2\right)^2}{ \cF_{40}^2
		\cF_{50}^2}+\frac{24 \cF_{30}^2}{\cF_{20}^2}\right) +O(e^{-3r}), &\\
	f_4(r) = &\cF_{40}e^{r} + \frac{e^{-r}}{240\cF_{40}}\left(-\frac{2  \left(\cF_{30}^2+13 \cF_{40}^2+\cF_{50}^2\right)
		\cF_{40}^2}{\cF_{20}^4}\right.&\\
	&\left.	+\frac{11 \cF_{30}^4+2 \left(\cF_{40}^2-11
		\cF_{50}^2\right) \cF_{30}^2-13 \cF_{40}^4+11 \cF_{50}^4+2
		\cF_{40}^2 \cF_{50}^2}{ \cF_{30}^2 \cF_{50}^2}+\frac{24
		\cF_{40}^2}{\cF_{20}^2}\right) + O(e^{-3r}), &\\
	f_5(r) =  &\cF_{50}e^{r} + \frac{e^{-r}}{240\cF_{50}}\left(-\frac{2  \left(\cF_{30}^2+\cF_{40}^2+13 \cF_{50}^2\right)
		\cF_{50}^2}{\cF_{20}^4}\right.&\\
	&\left.	+\frac{11 \cF_{30}^4+\left(2 \cF_{50}^2-22
		\cF_{40}^2\right) \cF_{30}^2+11 \cF_{40}^4-13 \cF_{50}^4+2
		\cF_{40}^2 \cF_{50}^2}{ \cF_{30}^2 \cF_{40}^2}+\frac{24
		\cF_{50}^2}{\cF_{20}^2}\right) + O(e^{-3r}).&\\
\end{aligned}
\ee
    Same as argued after \eqref{UVExpansionAnsatz2}, we expect the coefficients $\{ \cF_{j7}\}^{j = 2,3,4}$ to vanish, thus the family of solutions have four free parameters.
    
\subsection{Small radius expansion and numerics - NUT}
\begin{figure}[ht]
	\includegraphics[width=9cm]{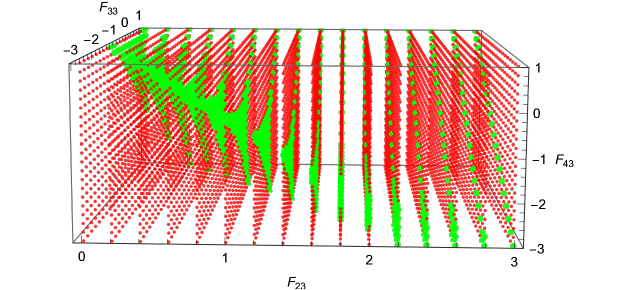}\quad 
	\includegraphics[width=7cm]{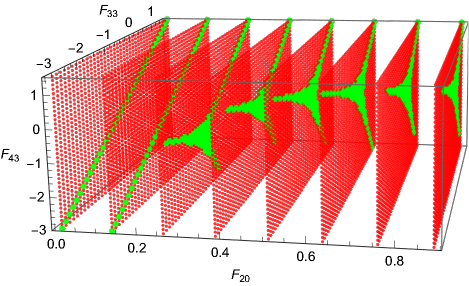}
	\caption{\rm Left: Green points correspond to initial values for NUT solution that integrate to infinity, and red points for those not. Right: Same plot for the Bolt solution.}
	\label{UVAllowedPoints}
\end{figure}

For NUT, we assume the following small-$r$ expansion:
\be
\begin{aligned}
	&f_1(r) = 1, &\\
	&f_i(r) = F_{ij}(r-r_0)^j,\quad i = 2,3,4,5,\quad j = 1,2,3,\cdots. &\\
\end{aligned}
\ee
In addition, we assume the SO(5)$\times$SO(3) symmetry to be preserved along the constant-$r$ surface, thus we require
\be
F_{31} = F_{41} = F_{51} > 0.
\ee
The equation of first order has two solutions:
\be
\left\{
\begin{aligned}
	&F_{21} = \frac{1}{2},&\\
	&F_{31} = F_{41} = F_{51} = \frac{1}{2},&
\end{aligned}
\right.\qquad {\rm or}\qquad
\left\{
\begin{aligned}
	&F_{21} = \frac{3\sqrt{5}}{10},&\\
	&F_{31} = F_{41} = F_{51} = \frac{3}{10}.&
\end{aligned}
\right.
\label{twoChoices}
\ee
For the first choice, we have solved \eqref{EOMAnsatz1} up to order $O(r-r_0)^{13}$, there're three free parameters, which we can choose to be $F_{23}, F_{33},F_{43}$, the first several terms up to $O(r-r_0)^5$ are:
\be
\begin{aligned}
	f_2(r) = &\frac{\rho }{2}+\rho ^3 F_{23}+\rho ^5 \left[-\frac{2}{5}  \left(F_{33}^2+F_{43} F_{33}+F_{43}^2\right)+\frac{7}{30}   (F_{33}+F_{43})\right.&\\
	&\left.	-\frac{1}{60} F_{23} (96  
	(F_{33}+F_{43})-91)-9 F_{23}^2-\frac{49}{720}\right]+O\left(\rho ^7\right), &\\
	f_3(r) = &\frac{\rho }{2  }+\rho ^3 F_{33}-\frac{\rho^5}{720  } \left[-144  \left(13 F_{33}^2-12 F_{43} F_{33}-12 F_{43}^2\right)+252   (F_{33}-4 F_{43})\right.&\\
	&	\left.+96 F_{23} (12   (F_{33}+6
	F_{43})-7)+49\right]+O\left(\rho ^7\right), &\\
	f_4(r) = &\frac{\rho }{2  }+\rho ^3 F_{43}-\frac{\rho^5}{720  } \left[44  \left(12 F_{33}^2+12 F_{43} F_{33}-13 F_{43}^2\right)-252   (4 F_{33}-F_{43})\right.&\\
	&\left.+96 F_{23} (12   (6 F_{33}+F_{43})-7)+49\right]+O\left(\rho ^7\right), &\\
	f_5(r) = &\frac{\rho }{2  }+\frac{\rho ^3}{12  } (-12   (F_{33}+F_{43})-48 F_{23}+7) + \frac{\rho ^5}{240  } \left[48  \left(13 F_{33}^2+38 F_{43} F_{33}+13 F_{43}^2\right)\right. \\
	&\left. -644(F_{33}+F_{43}) + 112 F_{23} (48   (F_{33}+F_{43})-23)+11520 F_{23}^2+147\right]+O\left(\rho ^7\right),&\\
\end{aligned}
\ee
where $\rho \equiv r - r_0$. When assigning $F_{23} = F_{33} = F_{43} = 1/12$, we obtain the standard ${\rm AdS_8}$ solution with $f_2 = f_3 = f_4 = f_5 = \sinh \rho$. Take the small $r$ expansion as the initial condition, we can perform numerics for general value of $\{ F_{23}, F_{33}, F_{43} \}$. As expected, not all initial values integrate to infinity: some of them vanishes at some finite $r$. By numerical simulation, we determine which initial values can be integrated to infinity and plot them in Fig.\ref{UVAllowedPoints}. There're two observations to be made. First, on each slice of the plot, the green region has the same shape as that obtained in \cite{Bobev:2016sap}. Second, initial values with $F_{33} = F_{43} = F_{53}$ are always allowed, which correspond to the first metric ansatz.

For the second choice in (\ref{twoChoices}), up to order $O(r-r_0)^{13}$, the solution is fixed, with no free parameter, whose leading order expansions can be identified with hyperbolic sine functions:
\be
\begin{aligned}
	&f_2(r) = \frac{3\sqrt{5}}{10}\left(\rho +\frac{\rho ^3}{6}+\frac{\rho ^5}{120}+\frac{\rho ^7}{5040}+\frac{\rho^9}{362880} + O\left(\rho ^{11}\right)\right) \rightarrow  \frac{3\sqrt{5}}{10}\sinh\rho, &\\
	&f_3(r)  = \frac{3}{10}\left(\rho +\frac{\rho ^3}{6}+\frac{\rho ^5}{120}+\frac{\rho ^7}{5040}+\frac{\rho ^9}{362880}+O\left(\rho^{11}\right)\right)\rightarrow \frac{3}{10}\sinh\rho, &\\	
	& f_3(r) = f_4(r) = f_5(r). &\\
\end{aligned}
\ee
With above solution, the metric has an asymptotic boundary given by squashed sphere in (\ref{SquashedSphereMetric1}) with $\lambda^2 = \frac{1}{5}$ as:
\be
ds^2 = dr^2 + \frac{9}{5} \sinh^2r ds_{\tilde{\mathbb{S}}^7}^2.
\ee
The curvatues of the geometry above is the same as the singular solution we obtained in the other metric (\ref{Curvatures}). 

\subsection{Small radius expansion and numerics - Bolt}
The ansatz for Bolt is
\be
\begin{aligned}
	&f_1(r) = 1, &\\
	&f_2(r) = F_{2j}(r-r_0)^j, j = 0,1,2,3,\cdots, &\\
	&f_i(r) = F_{ij}(r-r_0)^j, j = 1,2,3\cdots, &\\
	&F_{31} = F_{41} = F_{51} > 0,&\\
\end{aligned}
\ee
where we require $F_{20}>0$ for convenient. The leading terms are
\be
\begin{aligned}
	f_2(r) = & F_{20}-\frac{\rho ^2 \left(-7 F_{20}^2-3\right)}{8 F_{20}}-\frac{\rho ^4 \left(49 F_{20}^4+98 F_{20}^2+39\right)}{384 F_{20}^3}+\frac{\rho ^6}{46080 F_{20}^5} \\
	& \left[7 F_{20}^6 \left(1536  \left(F_{33}^2+F_{43}
	F_{33}+F_{43}^2\right)+1127\right)\right.\\
	&\quad  \left.	+3 F_{20}^4 \left(1536  \left(F_{33}^2+F_{43} F_{33}+F_{43}^2\right)+896   (F_{33}+F_{43})+5047\right)\right.\\
	&\quad \left.	+F_{20}^2 (1152  
	(F_{33}+F_{43})+10465)+2475\right]+O\left(\rho ^8\right). \\
	f_3(r) = & \frac{\rho }{2  }+\rho ^3 F_{33}+\frac{\rho ^5}{480   F_{20}^2} \left(F_{20}^2 \left(576  \left(2 F_{33}^2-3 F_{43} F_{33}-3 F_{43}^2\right)-420   F_{33}+49\right)\right.\\
	&\quad \left.	-72   (F_{33}+6
	F_{43})+35\right)+O\left(\rho ^7\right). \\
	f_4(r) = & \frac{\rho }{2  }+\rho ^3 F_{43}+\frac{\rho ^5}{480   F_{20}^2} \left(F_{20}^2 \left(576 \left(2 F_{43}^2 - 3 F_{43} F_{33}-3 F_{33}^2\right)-420   F_{43}+49\right) \right.\\
	& \quad \left. -72   (6F_{33}+F_{43})+35\right)+O\left(\rho ^7\right). \\
	f_5(r) = & \frac{\rho }{2  }+\rho ^3 \left(-\frac{1}{4   F_{20}^2}-F_{33}-F_{43}\right)\\
	&\quad +\frac{\rho ^5}{480   F_{20}^4} \left(F_{20}^4 \left(576  \left(2 F_{33}^2+7 F_{43} F_{33}+2 F_{43}^2\right)+420  
	(F_{33}+F_{43})+49\right)\right.\\
	&\quad \left.	+4 F_{20}^2 (162   (F_{33}+F_{43})+35)+90\right)+O\left(\rho ^7\right).\\
\end{aligned}
\ee
Taking the first two orders as the initial conditions, one can also solve the Einstein equations (\ref{EOMAnsatz1}) numerically. We can also obtain the set of initial values which can be integrated to infinity, as plotted in Fig.\ref{UVAllowedPoints}. The shape of the region is also similar to the one obtained in \cite{Bobev:2016sap}.


\section{Conformal mapping from sphere to plane}
\label{ConformalMapping}
In the main text, to perform the integration of the stress tensor correlation function, one needs the explicit conforming map from $\mathbb{S}^7$ to $\mathbb{R}^7$. In this section, we'll state how we get it and discuss how to construct the map in different situations.

Generally speaking, the conformal map is the composition of an embedding map from $\mathbb{S}^d$ to $\mathbb{R}^{d+1}$, together with a standard stereographic projection from $\mathbb{R}^{d+1}$ to the equatorial plane $\mathbb{R}^d$:
\be 
Y^{\bar a} = \frac{y^{\bar a} }{1-y^{d+1}},\quad \bar{a} = 1,2,...,d.\quad Y^{\bar{a}} \in \mathbb{R}^d,\quad y^A \in \mathbb{R}^{d+1}.
\label{StereographicProjection}
\ee 
The conformal factor is given by
\be 
ds^2_{\mathbb{R}^{d}} = \Omega^2 ds^2_{\mathbb{R}^{d+1}}  = \Omega^2 ds^2_{\mathbb{S}^d},\quad \Omega = \frac{1}{2}(1+Y_{\bar{a}} Y^{\bar{a}} ) = \frac{1}{1-y^{d+1}}.
\ee 
Thus the only step dependent on the specific metric is the embedding map from $\mathbb{S}^d$ to $\mathbb{R}^{d+1}$, which we'll discuss. In general, the exercise of obtaining the embedding map from the induced coordinate is a highly non-trivial task even for two-dimensional cases. The examples considered in this section are only some special cases where such a map can be constructed systematically.

\subsection{U(1) bundle}
\label{U1Bundle}

We start with an easier and more established situation, i.e., U(1) Hopf fibration. Spaces constructed by Hopf fibration with U(1) bundle are common in the literature, and closely related to generalized Taub-NUT spaces. One of them is $\mathbb{S}^{2k+1}$ constructed by U(1) bundle over $\mathbb{CP}^k$, preserving U(1)$\times$ SU$(k+1)$ symmetry. The metric can be written explicitly:
\be 
ds^2_{\mathbb{S}^d} = \frac{1}{d+1}ds^2_{\mathbb{CP}^k} + \left( d\psi + \frac{A_{\mathbb{CP}^k}}{d+1} \right)^2,\quad d = 2k+1,\ 0\le \psi\le 2\pi,
\label{U1FiberSphere}
\ee
 where the Fubini-Study metric on the projective space and the K\"ahler potential $A_{\mathbb{CP}^k}$ are defined recursively:
\be \begin{aligned}
	ds^2_{\mathbb{CP}^k} &= (2k+2)\left[ d\xi_k^2 + \frac{1}{2k} \sin^2\xi_k ds^2_{\mathbb{CP}^{k-1}} +  \sin^2\xi_k \cos^2\xi_k \left( d\psi_k + \frac{1}{2k} A_{\mathbb{CP}^{k-1}} \right)^2 \right], \\
	A_{\mathbb{CP}^k} &= (2k+2)\sin^2\xi_k\left( d\psi_k + \frac{1}{2k} A_{\mathbb{CP}^{k-1}} \right),\quad 0\le \xi_i\le \frac{\pi}{2},\ 0\le\psi_i\le 2\pi.  \\
\end{aligned}
\ee 
The lowest value is $k = 1$, which reproduces the initial construction of Hopf fibration for $\mathbb{S}^3$:
\be 
ds^2_{\mathbb{CP}^1} = 4(d\xi_1^2 + \sin^2\xi_1 \cos^2\xi_1 d\psi_1^2),\quad A_{\mathbb{CP}^1} = 4\sin^2\xi_1 d\psi_1.
\ee 
The metric on $\mathbb{S}^3$ is:
\be 
ds^2 = d\xi_1^2 + \cos^2\xi_1\sin^2\xi_1 d\psi_1^2 + \left( d\psi + \sin^2\xi_1 d\psi_1 \right)^2 .
\ee 
To make the SO(4) invariance manifest, we can do the following trick: replace $d\psi^2$ by $(\cos^2\xi_1 + \sin^2\xi_1)d\psi^2$, collect terms with $\sin^2\xi_1$ and $\cos^2\xi_1$ respectively, then the metric becomes:
\be 
ds^2 = d\xi_1^2 + \cos^2\xi_1 d\psi^2 + \sin^2\xi_1 (d\psi + d\psi_1)^2 .
\label{StdMetricS3}
\ee 
This form may look more familiar, which can be obtained by the simple embedding formula, as we'll explain later:\footnote{Here we reverse the order of $x^a$ for the convenience of evaluating the integral.}
\be 
x^4 = \cos\xi_1 \cos \psi,\quad x^3 = \cos\xi_1\sin\psi,\quad x^2 = \sin\xi_1 \cos (\psi+\psi_1),\quad x^1 =  \sin\xi_1 \sin (\psi+\psi_1).
\label{EmbeddingMap}
\ee 
Now we head forward to $k=2$ for a metric on $\mathbb{S}^5$. Again, this time we replace $d\psi^2$ by $(\cos^2\xi_2 + \sin^2\xi_2)d\psi^2$, leave the former on its own, and put the latter together with the other terms:
\be 
ds^2 = d\xi_2^2 + \cos^2\xi_2 d\psi^2 + \sin^2\xi_2 \left[ d\xi_1^2 + \cos^2\xi_1 (d\psi + d\psi_2)^2 + \sin^2\xi_1 (d\psi + d\psi_1 + d\psi_2)^2 \right],
\label{StdMetricS5}
\ee 
where the term in the bracket is organized in order to produce the same form as (\ref{StdMetricS3}). So we can write down the embedding formula easily:
\be \begin{aligned}
	& x^6 = \cos\xi_2 \cos\psi,\quad x^5 = \cos\xi_2\sin\psi, \\
	& x^4 = \sin\xi_2 \cos \xi_1 \cos(\psi+\psi_2),\quad x^3 = \sin\xi_2 \cos \xi_1 \sin (\psi+\psi_2), \\
	&x^2 = \sin\xi_2 \sin \xi_1 \cos (\psi + \psi_1 + \psi_2),\quad x^1 = \sin\xi_2 \sin \xi_1 \sin (\psi + \psi_1 + \psi_2) .
\end{aligned}\ee 
Now we understand how the procedure works. Let's take one last example, the metric on $\mathbb{S}^7$ with $k=3$. Again, we replace $d\psi^2$ by $(\cos^2\xi_3 + \sin^2\xi_3)d\psi^2$ first:
\be 
ds^2 = d\xi_3^2 + \cos^2\xi_3 d\psi^2 + \sin^2 \xi_3 [\ A\ ].
\ee 
In our expectation, we need to organize $A$ in a form similar to (\ref{StdMetricS5}), where we need to depart $d\psi^2$ again by $(\cos^2\xi_2 + \sin^2\xi_2)d\psi^2$, finally we get
\be \begin{aligned}
	A & = d\xi_2^2 + \cos^2\xi_2 (d\psi + d\psi_3)^2 \\
	& + \sin^2\xi_2 \left[ d\xi_1^2 + \cos^2\xi_1 (d\psi + d\psi_2 + d\psi_3)^2 + \sin^2\xi_1 (d\psi + d\psi_1 + d\psi_2 + d\psi_3)^2 \right] .
\end{aligned}
\ee 
We can then write down the embedding map without any difficulty:
\be \begin{aligned}
	&x^8 = \cos\xi_3 \cos\psi,\quad x^7 = \cos\xi_3 \sin\psi,\quad x^6 = \sin\xi_3 \cos\xi_2 \cos(\psi + \psi_3), \\
	&x^5 = \sin\xi_3 \cos\xi_2 \sin(\psi + \psi_3),\quad x^4 = \sin\xi_3 \sin\xi_2 \cos\xi_1 \cos (\psi + \psi_2 + \psi_3), \\
	& x^3 = \sin\xi_3 \sin\xi_2 \cos\xi_1 \sin (\psi + \psi_2 + \psi_3),\quad x^2 = \sin\xi_3 \sin\xi_2 \sin\xi_1 \cos(\psi + \psi_1 + \psi_2 + \psi_3), \\
	& x^1 = \sin\xi_3 \sin\xi_2 \sin\xi_1 \sin(\psi + \psi_1 + \psi_2 + \psi_3).
\end{aligned}\label{S7Embedding}\ee 

Upon following the procedures above, we can write down the mapping for general $k$:
\be\begin{aligned}
	& x^{2k+2} = \cos\xi_k \cos\psi ,\quad x^{2k+1} = \cos\xi_k \sin\psi, \\
	& x^{2k} = \sin\xi_k \cos\xi_{k-1} \cos\theta_k,\quad x^{2k-1} = \sin\xi_k \cos\xi_{k-1} \sin\theta_k, \\
	& x^{2k-2} = \sin\xi_k \sin\xi_{k-1} \cos\xi_{k-2} \cos\theta_{k-1},\quad x^{2k-3} = \sin\xi_k \sin\xi_{k-1} \cos\xi_{k-2} \sin\theta_{k-1}, \\
	&... ... \\
	& x^4 = \prod_{i=2}^k \sin\xi_i \cos\xi_1 \cos\theta_2,\quad x^3 =  \prod_{i=2}^k \sin\xi_i \cos\xi_1 \sin\theta_2, \\
	& x^2 =  \prod_{i=1}^k \sin\xi_i \cos\theta_1,\quad x^1 =  \prod_{i=1}^k \sin\xi_i \sin\theta_1,
\end{aligned}\label{GeneralEmbedding}\ee 
where we have defined
\be 
\theta_i = \psi + \sum_{j = i}^k \psi_j,\quad i = 1,2,...,k.
\ee 
With the embedding map, one can write down the conformal map from spheres to planes, and evaluate the integrated two-point function of the stress tensor (\ref{Integrated2ptFunc}). We have evaluated this explicitly, and find it corresponds to the prediction of (\ref{Prediction}).

\subsection{SU(2) bundle}
Our metric on $\mathbb{S}^7$ that preserves explicitly SO(5)$\times $SO(3) isometry is the simplest example of Hopf fibration with SU(2) bundle, thus we expect some tricks can help us obtain the embedding map like (\ref{EmbeddingMap}), which is the simplest fibration with U(1) bundle. We discussed in section \ref{ATableTwoMetrics} that there're two equivalent ways to construct metrics on a sphere: by embedding into projective space, or Hopf fiber over projective space. As we will see, the embedding map of the first class of metrics can be obtained by the identification of projective space with Euclidean space in the near-origin limit. And the embedding map of the second class of metrics can be obtained by an explicit coordinate transformation \eqref{relation} from the first class.

Let's start with the first class of metrics. To illustrate how it works, we take an example first and consider $\mathbb{S}^3 \in \mathbb{CP}^2$, following the same steps in section \ref{ATableTwoMetrics}, we can obtain the metric on $\mathbb{S}^3$. The Fubini-Study metric is given by
\be 
ds^2 = (1+\bar{q}_kq_k)^{-1} d\bar{q}_i dq_i - (1+\bar{q}_kq_k)^{-2}\bar{q}_idq_id\bar{q}_jq_j,\quad q_1,q_2\in \mathbb{CP}^2.
\ee
We take the following parametrization on $\mathbb{CP}^2$:
\be\begin{aligned} 
	q_1 &= U\tan\chi\cos\frac{\mu}{2},\quad q_2 = V\tan\chi\sin\frac{\mu}{2},\quad U = e^{\mathbf{i}\theta/2},\quad V = e^{\mathbf{i}\Theta/2}, \\
	& \quad 0\le \mu \le \pi,\ 0\le \chi \le \frac{\pi}{2},\ 0\le \theta, \Theta \le 4\pi.
\end{aligned}\label{ParaCP2}\ee 
The Maurer-Cartan form and left-invariant one-form are
\be 
2U^{-1}dU = \mathbf{i}\sigma,\quad 2V^{-1}dV = \mathbf{i}\Sigma ,\quad \sigma = d\theta,\quad \Sigma = d\Theta.
\ee 
The Fubini-Study metric on $\mathbb{CP}^2$ is given by
\be 
\begin{aligned}
	ds^2 &= d\chi^2 + \frac{1}{4}\sin^2\chi \left[ d\mu^2 + \frac{1}{4}\sin^2\mu \omega^2 + \frac{1}{4}\cos^2\chi (\nu + \omega \cos\mu)^2 \right], \\
	& \quad \nu \equiv \sigma+\Sigma,\quad \omega \equiv \sigma - \Sigma .
\end{aligned}
\ee 
The line element in the bracket is the squashed three-sphere metric, with the squashing parameter identified with $\lambda\equiv \cos\chi$. In the limit $\chi\rightarrow 0$, we get the metric on the round sphere:
\be 
ds^2_{\mathbb{S}^3} = \frac{1}{4}\left( d\mu^2 + \cos^2 \frac{\mu}{2}d\theta^2 + \sin^2\frac{\mu}{2}d\Theta^2\right) .
\ee 
To compare with (\ref{StdMetricS3}), we can define $\tilde{\mu} \equiv \mu/2, \tilde{\theta}\equiv \theta/2,\tilde{\Theta}\equiv\Theta/2$ to get rid of the $1/4$ factor:
\be 
ds^2 = d\tilde{\mu}^2 + \cos^2\tilde{\mu}d\tilde{\theta}^2 + \sin^2\tilde{\mu}d\tilde{\Theta}^2.
\label{MostStdMetricS3}
\ee 
The metric (\ref{StdMetricS3}) is related to this one by a twist over angles:
\be 
\tilde{\mu} = \xi_1,\quad \tilde{\theta} = \psi,\quad \tilde{\Theta} = \psi + \psi_1 .
\ee 
Now we're ready to discuss the embedding map of $\mathbb{S}^3$ in $\mathbb{R}^4$. It still seems non-trivial to write down an embedding map from (\ref{MostStdMetricS3}) to $\mathbb{R}^4$, but the magic happens at $\chi\rightarrow 0$ limit of $\mathbb{CP}^2$: the space is identical to $\mathbb{R}^4$. We write down the map between $\mathbb{CP}^2$ and $\mathbb{R}^4$:
\be 
q_1 = x^1 + \mathbf{i}x^2,\quad q_2 = x^3 + \mathbf{i} x^4 .
\ee     
Compare it with (\ref{ParaCP2}), one obtains the following map between $(\mu, \theta, \Theta)$ and $(x^1,x^2,x^3,x^4)$:
\be 
x^1 = \cos\frac{\mu}{2}\cos\frac{\theta}{2},\quad x^2 = \cos\frac{\mu}{2}\sin\frac{\theta}{2},\quad x^3 = \cos\frac{\mu}{2}\cos\frac{\Theta}{2},\quad x^4 = \cos\frac{\mu}{2}\sin\frac{\Theta}{2}.
\ee 
This provides an embedding rule from the round sphere metric of the first class. However, the metrics that we're more interested in belong to the other class, for example, those we discuss in the last section: they're all obtained by Hopf fiber. For U(1) bundle, the problem is not severe as we can identify the two metrics (\ref{StdMetricS3}) and (\ref{MostStdMetricS3}) by observation. But for SU(2) bundle, it's no longer direct. Luckily, the map (\ref{relation}) between them has been worked out forty years ago. \cite{Awada:1982pk, Duff:1986hr}

The example we discuss in the main text is $\mathbb{S}^7$ embedded in $\mathbb{HP}^2$. Similarly, we identify the near-origin limit of $\mathbb{HP}^2$ as $\mathbb{R}^8$:
\be 
q_1 = x^1 + x^2 \mathbf{i} + x^3 \mathbf{j} + x^4 \mathbf{k},\quad q_2 = x^5 + x^6 \mathbf{i} + x^7 \mathbf{j} + x^8 \mathbf{k}.
\ee 
Meanwhile,
\be     
q_1 = \cos\frac{\mu}{2} \tilde{U},\quad q_2 = \sin\frac{\mu}{2} \tilde{V},
\ee 
where $\tilde{U}$ and $\tilde{V}$ are defined in terms of $(\mu, \Theta, \Phi, \Psi, \theta, \phi, \psi)$ according to (\ref{Utilde}). We get the following embedding map:
\be\small \label{EmbeddingMapS7}
\begin{aligned}
	&x^1 = \cos \frac{\mu }{2} \left(  \cos \frac{\theta }{2} \cos \frac{\Theta
	}{2} \cos \frac{1}{2} (-\Phi +\psi -\Psi +\phi )+\sin \frac{\theta }{2}
	\sin \frac{\Theta }{2} \cos \frac{1}{2} (-\Phi -\psi +\Psi +\phi ) \right), \\
	&x^2 =\cos \frac{\mu }{2} \left( \sin \frac{\theta }{2} \cos \frac{\Theta
	}{2} \cos \frac{1}{2} (\Phi -\psi +\Psi +\phi )-\cos \frac{\theta }{2}
	\sin \frac{\Theta }{2} \cos \frac{1}{2} (\Phi +\psi -\Psi +\phi )\right), \\
	&x^3 = \cos \frac{\mu }{2} \left( \sin \frac{\theta }{2} \cos \frac{\Theta
	}{2} \sin \frac{1}{2} (\Phi -\psi +\Psi +\phi )-\cos \frac{\theta }{2}
	\sin \frac{\Theta }{2} \sin \frac{1}{2} (\Phi +\psi -\Psi +\phi )\right), \\
	&x^4 = \cos \frac{\mu }{2}\left(  \sin \frac{\theta }{2} \sin \frac{\Theta
	}{2} \sin \frac{1}{2} (-\Phi -\psi +\Psi +\phi )+\cos \frac{\theta }{2}
	\cos \frac{\Theta }{2} \sin \frac{1}{2} (-\Phi +\psi -\Psi +\phi )\right), \\
	& x^5 = \cos \frac{\theta }{2} \sin \frac{\mu }{2} \cos \frac{\psi +\phi }{2}, \qquad\qquad x^6 = \sin \frac{\theta }{2} \sin \frac{\mu }{2} \cos \frac{\phi -\psi }{2}, \\
	& x^7 = \sin \frac{\theta }{2} \sin \frac{\mu }{2} \sin \frac{\phi -\psi }{2},\qquad\qquad  x^8 = \cos \frac{\theta }{2} \sin \frac{\mu }{2} \sin \frac{\psi +\phi }{2}. \\
\end{aligned}\ee 
The conformal map between $\mathbb{S}^7$ and $\mathbb{R}^7$ is obtained by combining the stereographic projection and the embedding map above, then we can evaluate the integral of the stress tensor two-point function as in the main text.

\section{Integrated correlators on general U(1) bundle spheres}
\label{FppForGeneralk}
In this section, we provide more details of our reproduction of the formula (\ref{Prediction}) for general $k$, where the integral (\ref{Integrated2ptFunc}) is performed on $\mathbb{S}^{2k+1}$, which we rewrite as below:
\be 
I = -\frac{C_T}{4}V_{2k+1} \int_{\mathbb{S}^{2k+1}} \sqrt{g_{2k+1}^{(0)}} d^{2k+1}x \Omega^{2k-1}(0)\Omega^{2k-1}(x) M^{\bar{a}\bar{b}}(0) M^{\bar{c}\bar{d}}(x) \frac{I_{\bar{a}\bar{b};\bar{c}\bar{d}}(X)}{X^{2(2k+1)}}
\ee 
The volume of $\mathbb{S}^{2k+1}$ and the square root of the metric (\ref{U1FiberSphere}) are given by:
\be 
V_{2k+1} = \frac{2\pi^{k+1}}{k!},\quad \sqrt{g_{2k+1}^{(0)}} = \prod_{i=1}^k \cos\xi_i\sin^{2i-1}\xi_i 
\ee 
The conformal factor and $X^2$ are known from the standard stereographic projection (\ref{StereographicProjection}):
\be 
\Omega(x) = \frac{1}{1-\cos\psi \cos\xi_k},\ \Omega(0) = \frac{1}{2};\quad X^{2} = \frac{1+\cos\psi\cos\xi_k}{1-\cos\psi\cos\xi_k}
\ee 
We define the new quantities:
\be 
M^{\bar{a}\bar{b}} = h^{ab}\frac{\partial X^{\bar{a}} }{\partial x^a}\frac{\partial X^{\bar{b}} }{\partial x^b} = -\frac{\partial X^{\bar{a}} }{\partial  \psi}\frac{\partial X^{\bar{b}} }{\partial \psi} \equiv -\Lambda^{\bar{a}} \Lambda^{\bar{b}},
\ee 
where we used the fact that the only non-vanishing component of $h^{ab}$ is $h^{\psi\psi} = -1$. Using another fact that the only non-zero component of $M^{\bar{a}\bar{b}}(0)$ is $M^{\bar{\psi}\bar{\psi}} = -\frac{1}{4}$, the contraction between $ M^{\bar{a}\bar{b}}M^{\bar{c}\bar{d}}$ and $I_{\bar{a}\bar{b};\bar{c}\bar{d}}$ can be simplied as:
\be \begin{aligned}
	M^{\bar{a}\bar{b}}(0) M^{\bar{c}\bar{d}}(x) I_{\bar{a}\bar{b};\bar{c}\bar{d}}(X) & = \frac{1}{4}(\Lambda^{2k+1})^2 - \frac{1}{X^2} (\Lambda^{\bar{a}} X_{\bar{a}}) (\Lambda^{2k+1}X_{2k+1})\\
	& + \frac{1}{X^4}(\Lambda^{\bar{a}} X_{\bar{a}})^2 (X^{2k+1})^2 - \frac{1}{4(2k+1)}(\Lambda^{\bar{a}}\Lambda^{\bar{a}})\\
\end{aligned}\ee 
Using our general form of coordinate transformation (\ref{GeneralEmbedding}), we can get a general form of the quantities appearing above, which, if we put together into the integrand together with what we have above, we obtain:
\be\begin{aligned} 
	F''(0) &= -\frac{C_T\pi^{k+1}}{(2k+1)2^{2k+5}k!} \int\sqrt{g}d^{2k+1}x \frac{1}{(1+\cos\psi\cos\xi_k)^{2k+3}}\left[ 8 k \cos 2 \psi \cos ^2\xi_k \right. \\
	&\left. \qquad  +4 \cos \psi[ (6 k-1) \cos \xi_k+(2 k+1) \cos
	3 \xi _k] +4 (3 k+1) \cos2 \xi _k \right. \\
	&\left. \qquad +(2 k+1) \cos 4 \xi _k+10 k-5 \right] 
\end{aligned}\label{IntegralFpp}\ee 
Note that the integral doesn't depend on $\psi_i$, which reflects the symmetry of the manifold. As a first step, we can integrate out $\prod_{i=1}^k d\psi_i$ and $\prod_{j=1}^{k-1}d\xi_j$:
\be 
\int \prod_{i=1}^k\prod_{j=1}^{k-1} d\psi_id\xi_j\sqrt{g} = \frac{2\pi^k\cos\xi_k\sin^{2k-1}\xi_k}{(k-1)!}
\ee 
Since the dependence of $\psi$ is relatively simple in (\ref{IntegralFpp}), we integrate over $\psi$ first. The trick is to identify the integral range $\psi\in [\pi,2\pi]$ to $[0,\pi]$ by the invariance of the integrand under $\psi \rightarrow -\psi$ and periodicity $\psi \sim \psi + 2\pi$, and integrate on a new variable $z = \cos\psi$. This gives us with a divergent integral of $\xi_k$, whose value, after throwing away the diverging terms, is:
\be 
F''(0) = - C_T \frac{k\pi^{2k+3/2}\Gamma\left(-k-\frac{1}{2}\right)}{4^k (k-1)!}
\ee 
After identifying $2k+1 = d$ and simplification, the result is exactly the formula (\ref{Prediction}) predicted by \cite{Bueno:2018yzo} from the analysis of high-derivative gravity.

	\nocite{*}
	\bibliographystyle{unsrturl}
	\bibliography{citations}    
	
\end{document}